\documentclass[fleqn,usenatbib]{mnras}

\usepackage{newtxtext,newtxmath}
\usepackage[T1]{fontenc}
\usepackage{longtable}
\DeclareRobustCommand{\VAN}[3]{#2}
\let\VANthebibliography\thebibliography
\def\thebibliography{\DeclareRobustCommand{\VAN}[3]{##3}\VANthebibliography}
\usepackage{graphicx}	% Including figure files
\usepackage{amsmath}	% Advanced maths commands
\usepackage{multirow}
\usepackage{hyperref}
\makeatletter
\newlength{\abovecaptionskip}%
\makeatother
\usepackage{threeparttable}%
\usepackage{multicol}

\title[The Instability Strips of Cepheids]{Bridging theory and observations in stellar pulsations: The impact of convection and metallicity on the instability strips of Classical and Type-II Cepheids}

\author[M Deka et al.]{
Mami Deka$^{1}$\thanks{E-mail:mamideka8@gmail.com}, 
Earl P. Bellinger$^{2,3,4}$, % \thanks{earlbellinger@gmail.com}
Shashi M. Kanbur$^{5}$,  %\thanks{shashi.kanbur@oswego.edu} 
Sukanta Deb$^{1,6}$, %\thanks{E-mail: sukanta.deb@cottonuniversity.ac.in}
Anupam Bhardwaj$^{7}$, \newauthor
\hspace{1.3mm}Hugh Riley Randall$^{5}$, 
Selim Kalici$^{5}$, 
Susmita Das$^{8,9}$
\\
$^{1}$Department of Physics, Cotton University, Panbazar, Guwahati 781001, Assam, India\\
$^{2}$Department of Astronomy, Yale University, CT, USA\\
$^{3}$Max Planck Institute for Astrophysics, Garching, Germany\\
$^{4}$Stellar Astrophysics Centre, Aarhus, Denmark\\ 
$^{5}$Department of Physics, State University of New York Oswego, Oswego, NY 13126, USA\\
$^{6}$Space and Astronomy Research Center, Cotton University, Panbazar, Guwahati 781001, Assam, India\\
$^{7}$Inter-University Center for Astronomy and Astrophysics (IUCAA), Post Bag 4, Ganeshkhind, Pune 411 007, India\\
$^{8}$Konkoly Observatory, HUN-REN Research Centre for Astronomy and Earth Sciences, Konkoly-Thege Mikl\'os \'ut 15-17, H-1121, Budapest, Hungary\\
$^{9}$ CSFK, MTA Centre of Excellence, Budapest, Konkoly Thege Miklós út 15-17., H-1121, Hungary
}

\date{Accepted XXX. Received YYY; in original form ZZZ}

\pubyear{2024}

\begin{document}
\label{firstpage}
\pagerange{\pageref{firstpage}--\pageref{lastpage}}
\maketitle

%Abstract of the paper
\begin{abstract} 
The effect of metallicity on the theoretical and empirical period-luminosity (PL) relations of Cepheid variables is not well understood and remains a highly debated issue.
Here, we examine empirical colour-magnitude diagrams (CMDs) of Classical and Type-II Cepheids in the Magellanic Clouds and compare those with the theoretically predicted instability strip (IS) edges. We explore the effects of incorporating turbulent flux, turbulent pressure, and radiative cooling into the convection theory on the predicted IS at various metallicities using \textsc{MESA-RSP}. 
%We 
%Here, we examine colour-magnitude diagrams (CMDs) of Classical and Type-II Cepheids in the Magellanic Clouds to improve our understanding of the impact of stellar physics on the edges of the instability strip (IS). 
%We explore the effects of incorporating turbulent flux, turbulent pressure, and radiative cooling into the convection theory at multiple metallicities. 
We find that the edges become redder with the increasing complexity of convection physics incorporated in the fiducial convection sets, and are similarly shifted to the red with increasing metallicity. %; similarly, the IS edges are shifted towards redder colour with increasing metallicity. 
The inclusion of turbulent flux and pressure improves the agreement of the red edge of the IS, while their exclusion leads to better agreement with observations of the blue edge. 
About $90\%$ of observed stars are found to fall within the predicted bluest and reddest edges across the considered variations of turbulent convection parameters. 
%However, about 20\% of first-overtone Cepheids in the LMC lay outside the instability strips of the models. %, potentially highlighting a route toward improving our understanding of pulsation and convection. 
Furthermore, we identify and discuss discrepancies between theoretical and observed CMDs in the low effective temperature and high luminosity regions for stars with periods greater than $\sim 20$ days. 
These findings highlight the potential for calibrating the turbulent convection parameters in stellar pulsation models or the prediction of 
a new class of rare, long-period, `red Cepheids', thereby improving our understanding of Cepheids and their role in cosmological studies. 
\end{abstract}


%%%%%%%%%%%%%%%%%%%%%%%%%%%%%%%%%%%%%%%%%%%%%%%%%%

%%%%%%%%%%%%%%%%% BODY OF PAPER %%%%%%%%%%%%%%%%%%

\section{Introduction}

Pulsating stars serve as excellent astrophysical laboratories for testing theories of stellar structure, evolution, and pulsation \citep{cox80}. 
Radially pulsating stars such as Classical Cepheids (CCs), RR Lyraes (RRLs), and Type-II Cepheids (T2Cs) provide stringent constraints for these models by enabling a quantitative comparison of their observed and predicted pulsation properties. These stars are of additional interest as they obey well-defined Period-Luminosity (PL) relations, making them important distance indicators as well as proxies for studying galactic structures \citep[e.g.,][]{deka22a,bhuy23}. 

CCs are intermediate-mass evolved stars with $\sim 3-20~\rm{M}_{\sun}$ \citep{bono00,ande16,muse22,espi22}.
They are Population~I stars located in the instability strip (IS) of the Hertzsprung-Russell diagram (HRD). For stars within the IS, small departures from hydrostatic equilibrium are amplified because of the increase in the opacity of the gas located in the ionization zones of the outer envelope on compression \citep{cox74}. 
The pulsations are mainly driven by the kappa mechanism and 
reinforced by the gamma mechanism \citep{cate15}.

CCs are used as the primary calibrators for the first rung of the cosmic distance ladder \citep[][and references therein]{reis22} and play a crucial role in the determination of the local value of the Hubble constant ($H_{0}$). 
However, the ongoing discord between $H_{0}$ values derived from cosmic microwave background data 
\citep{plan20} and measurements based on CCs and Type Ia
supernovae (SNe) \citep{reis22} warrants a detailed investigation of possible systematics in the CC PL relations. For example, the addition of metallicity term to the CC PL 
relations is found to improve the accuracy of the distance ladder \citep{brev22, reis22, bhar23}. 
The effect of metallicity on the PL relation has been investigated using both observed
\citep{ripe20,reis21,roma22,brev22,tren23, bhar24} and theoretical data 
\citep{bono99,bono08,capu00,marc05,ande16, somm22}. 
The value of the coefficient in the metallicity term (also known as the $\gamma$ term in literature) in the empirical period-luminosity-metallicity (PLZ) relation is found to vary over a wide range from $-0.5$ to $0.0$~~mag $\rm dex^{-1}$ depending on wavelength from near-infrared to optical band \citep{uada01, reis21, ripe22, brev22, free23, tren23}. 
On the other hand, most of the aforementioned studies based on non-linear convection models have found $\gamma>0$, except \citet{ande16} who obtained $\gamma<0$ using linear models that incorporate the effect of rotation. 
The debate surrounding the sign of $\gamma$ term from the empirical and theoretical points of view thus remains unresolved and necessitates further investigations.

The slope and intercept of PL relation are strongly coupled with the slope and intercept of IS edges
\citep{sand08}. A steeper/shallower instability strip implies a steeper/shallower PL slope \citep{simo97}. To improve the accuracy of the PL relation, a precise estimation of IS edges is essential. Besides, the
study of the IS can offer valuable insights into the physical mechanisms of the stars residing within this region. Also, the effect of metallicity on the PL relation is strongly correlated with the effect of metallicity on the IS edges. 
The IS boundaries of CCs have been studied extensively using theoretical models \citep{bono00,fior02,ande16,somm22}. 
The works of \citet{bono00a}, \citet{fior02} and \citet{somm22} have found a strong dependence of metallicity on the topology of the IS based on the theoretical formulation as outlined by \citet{bono00a,bono00b}. 
However, \citet{ande16} found a weak dependence of metallicity on the IS topology based on linear non-adiabatic (LNA) pulsation models when including the effect of rotation in their models. 
Therefore, in conjunction with different chemical compositions, varying input physics applied in computing the models also influences the location of the IS edges, thereby impacting the corresponding PL relationship.
In this work, we carry out a  detailed study of the effect of turbulent convection theory (more specifically turbulent convection parameters) and metallicity on the IS morphology of CCs.

In addition to CCs, we also include T2Cs in this work to study the effect of metallicity 
and turbulent convection parameters on the topology of their IS. As the Population-II 
counterparts of CCs, they present us with a unique opportunity to improve our understanding 
of these effects in the old, low-mass, metal-poor stellar populations. Similar to CCs, they 
also obey a tight PL relation, enabling their use as distance indicators 
\citep{smol16,das21,bhar17,bhar22,das24}, particularly in regions where CCs are not 
abundant and RRLs are very faint, thereby underscoring their crucial role in the cosmic 
distance scale. Nevertheless, the evolutionary and pulsation characteristics of T2Cs have 
been relatively understudied compared to other radial pulsators, despite their useful role 
in the distance determinations within the Milky Way and other nearby/satellite galaxies \citep[e.g.,][]{bhar17,wiel22,sici24}. 

T2Cs are low-mass stars ($\sim 0.5-0.8~\rm{M_{\sun}}$) with periods between $1-50$~days, although there remains some uncertainty regarding the precise upper and lower period limits \citep{cate15,smol16}. 
These stars are categorized into three distinct classes based on their periods:  BL Herculis (BL~Her), with periods between $1-4$~days; W Virginis (W~Vir), with periods between $4-20$~days; and RV Tauris (RV Tau), with periods above $20$~days and correspond to different evolutionary stages. 
We exclude RV Tauris stars based on their periods due to the ambiguity surrounding their upper mass limit \citep{bodi19} and diversity in their progenitors—ranging from young-massive to old stars in binary systems \citep{mani18}. 
Henceforth, when we refer to T2Cs in this work, it encompasses only BL~Her and W~Vir stars.

The present study makes use of both observed and theoretical data of CCs and T2Cs. 
This combined study will not only give us observationally consistent IS, but 
also put constraints on the input physics involved. The observed data are taken from OGLE-IV \citep{sosz15,sosz18}. We generate the
theoretical data using 
Modules for Experiments in Stellar Astrophysics \citep[\textsc{MESA},][]{paxt10,paxt13,paxt15,paxt18,paxt19,jerm23}.  

We generate the theoretical data using the new Radial Stellar Pulsations \citep[RSP;][]{paxt19} module in \textsc{MESA} for modelling linear/non-linear radial pulsations exhibited by variable stars like CCs, T2Cs, and RRLs \citep{smol08,paxt19}. 
We carry out an LNA radial pulsation stability analysis to obtain the growth rates of radial pulsation modes, their linear periods, and the mean luminosity.
From the growth rates, we then determine the boundaries of the instability strip (IS). 

\textsc{MESA-RSP} enables us to investigate the impact of various turbulent convection parameters on the properties of the models. 
It implements the time-dependent turbulent convection model by \citet{kuhf86} following the stellar pulsation treatment from \citet{smol08}. 
It is mainly governed by eight free dimensionless parameters:
\begin{enumerate}
    \item[] $\alpha$: Mixing length of the convection
    \item[] $\alpha_m$: Eddy viscosity
    \item[] $\alpha_s$: Source of turbulence
    \item[] $\alpha_c$: Convective energy transport
    \item[] $\alpha_d$: Dissipation of turbulence
    \item[] $\alpha_p$: Turbulent pressure
    \item[] $\alpha_t$: Turbulent flux
    \item[] $\gamma_r$: Radiative losses of convective eddy energy. 
\end{enumerate}
Four sets of turbulent convection, labeled A, B, C, and D, are defined based on different combinations of these parameters \citep{paxt19}. Set A corresponds to the simplest convection model, Set B adds radiative cooling; Set C adds turbulent pressure and turbulent flux; and set D includes all these effects simultaneously \citep{paxt19}. 
The free parameters that enter the convective models are kept the same as provided in Table~3 and 4 of \citet{paxt19}. For convenience, the parameters are listed in Table~\ref{table:convective_set}.

\begin{table}
	\centering
	\caption{Turbulent convective parameter sets from \citet{paxt19}. Set A is the simplest one; 
 set B includes radiative cooling; set C includes turbulent pressure and turbulent flux, and set D includes all of them.}
	\label{table:convective_set}
	\begin{tabular}{lcccccr} % four columns, alignment for each
		\hline
        Parameters & Set A &Set B& Set C & Set D \\ \hline
        Mixing-length, $\alpha$ & $1.5$ & $1.5$ & $1.5$ & $1.5$ \\
        Eddy-viscous dissipation, $\alpha_{m}$ & $0.25$ & $0.50$ & $0.40$ & $0.70$ \\
        Turbulent source, $\alpha_{s}$ & $\frac{1}{2}\sqrt{\frac{2}{3}}$ & $\frac{1}{2}\sqrt{\frac{2}{3}}$ & $\frac{1}{2}\sqrt{\frac{2}{3}}$ & $\frac{1}{2}\sqrt{\frac{2}{3}}$ \\
        Convective flux, $\alpha_{c}$ & $\frac{1}{2}\sqrt{\frac{2}{3}}$ & $\frac{1}{2}\sqrt{\frac{2}{3}}$ & $\frac{1}{2}\sqrt{\frac{2}{3}}$ & $\frac{1}{2}\sqrt{\frac{2}{3}}$ \\
        Turbulent dissipation, $\alpha_{d}$  & $\frac{8}{3}\sqrt{\frac{2}{3}}$ & $\frac{8}{3}\sqrt{\frac{2}{3}}$& $\frac{8}{3}\sqrt{\frac{2}{3}}$ & $\frac{8}{3}\sqrt{\frac{2}{3}}$ \\
        Turbulent pressure, $\alpha_{p}$  & $0$ &  $0$ &  $\frac{2}{3}$ & $\frac{2}{3}$ \\
        Turbulent flux, $\alpha_{t}$  & $0$ &  $0$ & $0.01$ & $0.01$  \\
        Radiative cooling, $\gamma_{r}$  & $0$ &  $2\sqrt{3}$ & $0$ & $2\sqrt{3}$  \\
      \hline
	\end{tabular}
\end{table}

% $\gamma_{r}$ is to account for the
% energy exchange between convective elements and its surroundings through radiation. 
When turbulent pressure, turbulent flux, and radiative cooling are neglected (i.e., $\alpha_{p}=\alpha_{t}=\gamma_{r}=0$) and values of $\alpha_{s},\alpha_{c}$ and $\alpha_{d}$ are kept the same as in Table~\ref{table:convective_set}, the time-independent version of the \citet{kuhf86} model reduces to the standard mixing-length theory \citep[MLT;][]{bohm58}. The turbulent pressure ($\alpha_{p}$) and convective flux ($\alpha_{c}$) parameters were introduced by \citet{yeck98} which were not present 
in the original \citet{kuhf86} model. These values are set to $\alpha_{p}=2/3$ and $\alpha_{c}\equiv\alpha_{s}$  \citep{smol08}. Radiative cooling $\gamma_{r}$ is set
to $2/3$ following \citet{wuch98}. For further details on these parameters, we refer the interested reader to \citet{kuhf86,wuch98,yeck98,smol08}.

The location of red and blue edges as well as the width of the IS are highly 
influenced by the value of $\alpha,\alpha_{m}$ and $\alpha_{t}$. Generally, $\alpha_{p}$ is 
attributed to the excitation, and $\alpha_{m}$ and $\alpha_{c}$ to the 
damping of stellar oscillations. For example, \citet{yeck98} found the turbulent eddy 
viscosity as the primary agent for damping pulsations. In case of red giant variables, 
\citet{xion21} have found that their contributions to the excitation and damping effect 
change depending on the stellar parameters $M$, $L$ and $T_{\rm eff}$ and the frequency of 
the mode. However, it is important to note that the turbulent convection model \citep{kuhf86} integrated in \textsc{MESA-RSP} is not equivalent to that of 
\citet{yeck98} and \citet{xion21}, and that the assumptions and simplifications made in these models are different as well. Furthermore, the driving and damping of stellar pulsation cannot be exclusively attributed to the free parameters of the convection model. These might vary from model to model and a detailed study on a specific model is necessary to find
out the main agents for the damping and driving mechanism \citep[for details see][]{houd15}.  

In addition to the IS, the resulting pulsation growth rate, radial velocity and light curve shapes are also highly sensitive to these parameters \citep{paxt19}.
It must be noted that the fiducial turbulent convection parameters given by \citet{paxt19} are not calibrated to observations. 
%It is therefore important to find the optimized set of parameters to reproduce the observational characteristics. 
In a recent study, \citet{kova23} and \citet{kova24} thoroughly examined the impact of these turbulent convection parameters on the radial velocity and light curves of several RRLs and proposed revised values for certain parameters. 
The present study instead focuses on the impact of the fiducial turbulent convection parameters as outlined in \citet{paxt19} on the ensemble-based properties of the observed stars. 
We plan to extend this work in a future paper by considering further variations in these parameters individually for each star, aiming for a more rigorous comparison with observations.

The remainder of the paper is organized as follows. 
In Section~\ref{sec:data}, we discuss the theoretical and observed data.  
Section~\ref{sec:method} presents the methodology to carry out the analysis. 
We discuss the results of the analysis in Section~\ref{sec:results}. 
Finally, a summary and discussion of the implications of our study are presented in Section~\ref{sec:summary}.

\section{Data}
\label{sec:data}
\subsection{Theoretical Data}
\label{sec:th_data}
We have computed static models using the following input stellar parameters: mass ($M$), luminosity ($L$), effective temperature ($T_{\rm{eff}}$), metal fraction ($Z$) and hydrogen fraction ($X$). 
We have constructed two fine grids of pulsation models for CCs and T2Cs after a thorough study of the previous literature \citep{bono00a,bono00b,ande16,somm22,bhar20,das21}. 

Although the typical lower mass limit for CCs is reported to be $\sim 3\rm M_{\sun}$ in most literature \citep[e.g.,][]{pile22,kurb23}, our investigation reveals that a grid spanning a mass range of $(3-11) \rm{M_{\sun}}$ fails to reproduce observations corresponding to lower magnitudes/periods. Through a rough estimation, we find that the mass of a CC with a period of $0.8$ days (the lowest period observed for an FU CC from OGLE-IV data) is approximately $\sim 2.3 \rm M_{\sun}$, using the period-mass relation from \citet{turn96}. Following this, we adopt a wide range of stellar masses $(2-11~\rm{M_{\sun}})$ to closely match the observed properties. We also adopt an effective temperature range of $(4000-8000~K)$ in the CC grid.

%\textbf{For CCs, we adopt a wide range of stellar masses $(2-11~\rm \rm{M_{\sun}})$ and effective temperatures $(4000-8000~K)$ to match with observations.  
%The lowest period of $0.8$ day for a FU Cepheid from the observed data corresponds to a mass of  $\sim 2.3 \rm M_{\sun}$ following
%the mass-period relation as given in  \citet{turn96}. 

%An approximate lower mass of $~2M_{\odot}$ has been considered in the grid corresponding to  following the mass-period relation of \citet{turn96}. }
%Although $2 \rm{M_{\sun}}$ may appear too small to be a CC, it is important to note that our initial grid, as shown in Fig.~\ref{fig:input_grid}, %was constructed with a mass range of $(3-11~\rm \rm{M_{\sun}})$. However, this initial grid failed to reproduce the observed properties of stars in %the lower magnitude region of the CMD. In fact, we have found that
%the mass calculated for a Cepheid having period $0.8$ day (lowest period for a FU CC from the observed OGLE-IV  data) to be  $\sim 2.3 \rm %M_{\sun}$ using the mass-period 
%relation for CCs as given in \citet{turn96}. Therefore, we have extended the lower limit of CC mass to $2 \rm{M_{\sun}}$, and consequently %luminosity, by applying the mass-luminosity law in the original grid to match the observations.} 

The lower/upper limit of luminosity for a particular mass is set by decreasing/increasing the mass-luminosity relation by $0.25$ dex \citep{ande16}. 
Three different metallicities [Fe/H]~$=$~($0.00$, $-0.34$, $-0.75$) suitable for the Galaxy, Large Magellanic Cloud (LMC) and Small Magellanic Cloud (SMC), respectively have been adopted for the input grid following \citet{roma08,gier18}. 

Similarly, the masses of T2Cs are quoted to be in the range $(0.5-0.8)~\rm M_{\sun}$ in literature \citep{smol16,bhar20,das21}. However, we have considered a wider range of $(0.3-1.0)~\rm M_{\sun}$ in order to match with observations \citep[also see][]{das24}. 
Additionally, we have considered a range of luminosities and effective temperatures spanning $(10-1000)~\rm L_{\sun}$ and $(4000-8000)$~K, respectively. The upper limit for luminosity is slightly extended to account for a few very high-luminosity T2Cs observed in the data. 
T2Cs from different regions are found to have different metallicities \citep{welc12}. A few of them as obtained in previous literature are listed 
as follows. T2Cs from all the Galactic globular clusters are found to have metallicities (in dex): [Fe/H]~<~$-1.0$ \citep{clem01}; Galactic field region: $-1.0$~<~[Fe/H]~<~$0.0$ \citep{schi11} and for the Bulge: $-1.4$~<~[Fe/H]~<~$0.6$ \citep{harr84,wall02} and the majority of field T2Cs from the Galactic Halo are found to have [Fe/H]~>~$-0.9$. Hence, we adopt a wide range of metallicities spanning $-2.0$~<~[Fe/H]~<~$0.0$ for T2C input grid. 
Since our model-to-observation comparisons are exclusively focused on data from the LMC and SMC, this adopted range of [Fe/H] will be sufficient for the scope of this study.

Mass fractions of hydrogen $X$ and metals $Z$ are obtained from $[\rm{Fe/H}]=\log{(Z/X)}-\log{(Z/X)}_{\odot}$, where we adopt the solar mixture $\left(Z/X\right)_{\odot}=0.0181$ from \citet{aspl09}.
%the [Fe/H] values via 
%assuming scaled solar abundance following \citet{aspl09}. 
%\begin{align}
%[\rm{Fe/H}]=&\log{\left(\frac{Z}{X}\right)}-\log{\left(\frac{Z}{X}\right)}_{\odot}  \label{eq:met1}%\\
%%\Rightarrow \log{\left(\frac{Z}{X}\right)}=&[\rm{Fe/H}]+\log{\left(\frac{Z}{X}\right)}_{\odot}. 
%\end{align}

We assume the chemical evolution law \citep[e.g.,][]{peim74} 
\begin{align}
Y= & Y_{0}+\frac{\Delta Y}{\Delta Z}\; Z
%\Rightarrow Y=&0.2485+1.54Z
\end{align}
with a primordial helium abundance ${Y_{0}=0.2485}$ from \citet{hins13} and helium-to-metal enrichment ratio
${\Delta Y/\Delta Z=1.54}$ from \citet{aspl09}.
\iffalse
We know that
\begin{align}
X+Y+Z=&1 \\ \label{eq:met2}
\Rightarrow X+(0.2485+1.54Z)+Z=&1 \\
\Rightarrow 1+2.54\left(\frac{Z}{X}\right)=&\frac{1-0.2485}{X} \\
\Rightarrow X=&\frac{1+2.54\left(\frac{Z}{X}\right)}{1-0.2485} \\
\Rightarrow X=&\frac{1+2.54\left(\frac{Z}{X}\right)}{0.7515}.
\end{align}
Finally, the values of $Z$ are obtained using Equation~\ref{eq:met1}.
\fi
The obtained $X$ and $Z$ values corresponding to the  [Fe/H] values are provided in Table~\ref{table:feh}. 

Using the aforementioned $M,L$ and $T_{\rm eff}$ values, we have constructed two fine grids for both the classes of variables (shown in Fig.~\ref{fig:input_grid}). 
The grid is constructed using Sobol numbers, i.e., quasi-random low discrepancy numbers \citep{sobo67}. 
The advantage of using such a grid over a linear grid or a random grid is that Sobol numbers are uniformly distributed and fill the input parameter space more rapidly. 
%The linear grid needs a large number of models to fill up the parameter space, so it fills the parameter space very slowly. 
%On the other hand,  a random number grid tends to generate clumpy points at random, resulting in random gaps in the parameter space. 
For more details on the use of Sobol numbers to construct a grid, the readers are referred to \citet{bell16}. 
%Given the substantial computational resources required for the model calculations, it is worthwhile to use a compact grid covering a wider range of stellar parameters typical to the particular pulsating star. 
Final grids are constructed by mapping the compositions $Z$ and $X$ for each class with the Sobol grid ($M,L,T_{\rm eff}$). Table~\ref{table:grid} presents a summary of the input parameters of the constructed grids.

\begin{table}
	\centering
	\caption{Chemical compositions of the adopted models.}
	\label{table:feh}
	\begin{tabular}{lccccccr} % four columns, alignment for each
		\hline
        Variable & [Fe/H] & $Z$&$X$ \\ \hline
        CCs & $-0.75$ & $0.00240$ & $0.74541$ \\
        & $-0.34$ & $0.00609$ & $0.73603$ \\
        & $\hphantom{-}0.00$ & $0.01300$ & $0.71847$\\ \hline
        T2Cs & $-2.0$ & $0.00014$ & $0.75115$ \\
        & $-1.50$ & $0.00043$ & $0.75041$ \\
        & $-1.35$ & $0.00061$ & $0.74996$ \\
        & $-1.00$ & $0.00135$ & $0.74806$ \\
        & $-0.50$ & $0.00424$ & $0.74073$ \\
        & $-0.20$  & $0.00834$ &  $0.73032$ \\
        & $\hphantom{-}0.00$  & $0.01300$ & $0.71847$ \\
        
      \hline
	\end{tabular}
\end{table}

\begin{table*}
	\centering
	\caption{Stellar parameter range for CCs and T2Cs used in constructing the pulsation model grids. }
	\label{table:grid}
	\begin{tabular}{lccccccr} % four columns, alignment for each
		\hline
        Class of variables & $M/\rm{M_{\sun}}$ & $L/\rm{L_{\sun}}$ & $T_{\rm eff}$~(K) & [Fe/H] ~(dex) & Total number of models \\ \hline
        CCs & $2-11$ & $36-50000$ & $4000-8000$ & $(0.0,-0.34,-0.75)$ & $16839$ \\
        T2Cs & $0.3-1.0$ & $10-1000$ & $4000-8000$ & $(-2.0,-1.5,-1.0,-0.5,0.0,0.5)$ & $14112$ \\
      \hline
	\end{tabular}
\end{table*}

The LNA analysis is performed using the four convective sets in  \textsc{MESA-RSP} \citep{smol08,paxt19}. 
\textsc{MESA-RSP} treats radiative transfer in the diffusion approximation, which is adequate for classical pulsators such as CCs, RRLs, 
T2Cs, and $\delta$~Scuti stars \citep{smol08, aika08, das21, deka22b,kurb23, das24}. 

RSP solves the pulsation equations on a Lagrangian mesh. 
The mesh structure is divided into inner and outer zones with respect to a specified anchor temperature (see Fig.~2 of \citealt{paxt19}). 
Initially, we have kept the total number of Lagrangian mass cells $N_{\rm total}=200$ with $N_{\rm outer}=60$ cells above the anchor.  
However, if this configuration failed to create the initial model, then we have considered $N_{\rm total}=150$ and $N_{\rm outer}=40$.
The choice of numerical parameters: inner boundary temperature ($T_{\rm{inner}}=2\times10^{6}$~K) and anchor temperature ($T_{\rm{anchor}}=11,000$~K) are kept as outlined in \citet{paxt19} (for more details, see Appendix~\ref{sec:test}). 
These parameters are defined in such a way that we achieve a good resolution of the pulsation driving region. 
An example inlist used for the LNA computation is provided in \url{https://github.com/mami-deka/LNA_CMD}.

\subsection{Observational Data}
In order to compare the theoretical findings with the observed data, the observed optical $(V, I)$- band light curves of fundamental (FU), first-overtone (FO) mode CCs and T2Cs are taken from the OGLE-IV database \citep{sosz15, sosz18}. The light curves are available for both LMC and SMC. In the present analysis, well-sampled light curves with more than 30 data points are chosen both for CCs and T2Cs. The stars marked as uncertain in the database are also excluded. A concise overview of the 
observed data is outlined in Table~\ref{table:observed_data}.

\begin{table}
    \centering
    \caption{Stars used in the present study.}
    \label{table:observed_data}
    \begin{tabular}{lcc} % three columns
        \hline
        Class & Galaxy & Number of Stars \\ \hline
        FU CC & LMC & 2056 \\
        FO CC & LMC & 1349 \\
        FU CC & SMC & 2419 \\
        FO CC & SMC & 1423 \\
        T2C & LMC & 160 \\
        T2C & SMC & 34 \\ \hline
    \end{tabular}
\end{table}

\section{Method}
\label{sec:method}
\subsection{Theoretical Data Processing}
The theoretical stellar grids (discussed in Section~\ref{sec:data}) are used as input pulsation models for the LNA computation. From the computed models, we choose those models with 
a positive growth rate of the radial fundamental mode and linear periods between $0.8-200$~days for the FU CCs. Similarly for the FO CCs, a positive growth rate of the radial first overtone mode and inear periods less than $10$~days are chosen for further analysis. Likewise, T2Cs models are chosen from the models which have a positive growth rate of the radial fundamental mode and linear periods between $1-20$~days. The period range for theoretical models is chosen based on the observed periods. We use the pre-computed bolometric corrections (BCs) tables from \citet{leje98} to obtain the absolute magnitude of a model in the $V$ and $I$- bands.
Detailed conversion steps are given in \citet{paxt18} and briefly summarized in \citet{das21}.

\subsection{Observational Data Processing}
To compare theoretical prediction with observation in CMD, we require absolute
magnitudes and extinction corrected colour. 
First, we phase the light curves using the following equation:
\begin{align}
\label{eq:phase}
\Phi=&\frac{t-t_{0}}{P}-{\rm Int}\left(\frac{t-t_{0}}{P}\right),
\end{align}
where $t_{0}$, $P$ and $t$ represent the epoch of maximum light for the  $V$- band, the pulsation period 
of a star in days and the times of observations, respectively. 
These values are taken from the OGLE-IV database.

Then to obtain the mean magnitude from each light curve, we have fitted the phased light curves with a Fourier sine series \citep{deb09,deka22b} of the following form:
\begin{align}
\label{eq:Fourier}
m(t)=&A_{0}+\sum\limits_{i=1}^N A_{i} \sin\left(2\pi i \Phi+\phi_{i}\right),
\end{align}
where $A_{0}$ and $\omega=\frac{2\pi}{P}$ denote the mean magnitude and the angular frequency, respectively. The $i$th 
order Fourier coefficients are represented by  $A_{i}$ and $\phi_{i}$. The order of the fit $N$ is obtained using Baart's 
criteria \citep[and references therein]{deb09} by varying it between ($6-10$), ($6-10$), ($4-8$) for the $I$-band light 
curves of FU CCs, FO CCs and T2Cs respectively, while it is varied from $4-8$ for the corresponding $V$-band light curves.

The optical reddening map of \citet{skow21} for the Magellanic Clouds is used to obtain the extinction-corrected observed 
apparent mean magnitudes. The $E(V-I)$ reddening values obtained from \citet{skow21} map are converted into the extinctions
$A_{I}$ and $A_{V}$ using the relations $A_I=1.5\,E(V-I)$ and $A_{V}=2.5\,E(V-I)$ \citep{skow21}. The true apparent magnitudes are 
then converted into the corresponding absolute magnitudes assuming a mean distance modulus of  $\mu=18.48$ mag \citep{piet19} and
$\mu=18.977$ mag \citep{grac20} for the LMC and SMC, respectively.  

Furthermore, some of the observed data are discarded based on the approach following \citet{mado17} due to 
their higher horizontal deviations in the CMD. The CMDs in Fig.~\ref{fig:obs_cmd} highlight these deviated points in black colour. 
These deviated flares are presumably due to uncertainties associated with the magnitude and colour determinations and the reddening values adopted for each star. We employ a $2.5\sigma$ clipping to the linear fit on the magnitude residuals of the period-luminosity (PL) relation versus the corresponding residuals of the period-Wesenheit (PW) relation 
to remove the stars with high vertical dispersion.

% \begin{center}
% \input{star_nc.tex}
% \end{center}
\section{Results and discussion}
\label{sec:results}
\subsection{Instability strip edges}
\label{sec:results1}
We now aim to determine the IS blue and red edges using the theoretical data generated by \textsc{MESA-RSP}.
We use machine learning to train a linear classifier using the Support Vector Machine algorithm that we optimize using Stochastic Gradient Descent (SGD). 

% Evidence of non-linearity is also visible in the high magnitude region of our theoretical CMD; particularly at the red edge where $V$-band magnitude is greater than $\sim -4$~mag  (for example, see Fig.~\ref{fig:cmd_cep_FU_lmc_th}).
We note that while a non-linear classifier would be appropriate for some of the models considered, it is challenging to establish a simple relation coming out of such a classifier. Hence, we have enforced linear IS edges to provide simple relations that can be adopted for a direct comparison with observations.
We train the classifier on the theoretical data to distinguish models with negative and positive growth rates, and hence to find the boundaries of the IS.
We use the implementation from the scikit-learn python package \citep{pedr11}. 
Besides the determination of IS edge, we have also compared the theoretical data with the observed data in CMD to see how well our theoretical models are able to reproduce the observed properties. 

\begin{figure*}
\includegraphics[width=0.9\textwidth,keepaspectratio]{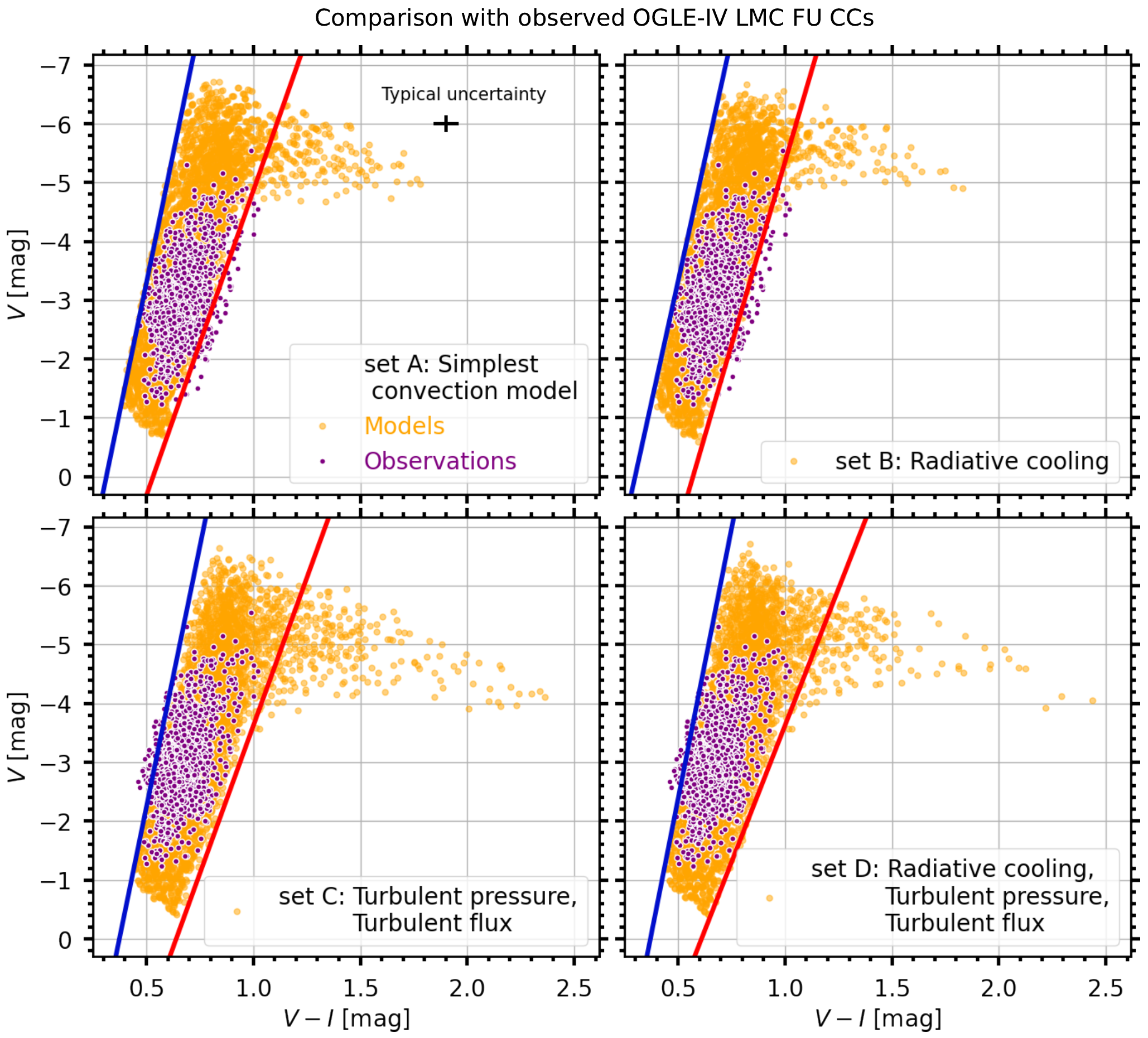}
\caption{Comparison of the color-magnitude diagram between observed fundamental-mode Cepheids in the LMC (purple dots) and theoretical models (orange dots) for four different prescriptions of convection. The theoretical blue and red edges from each convection set are shown with lines. %Blue and red lines represent the blue and red edges of the IS, respectively. 
The black cross in the top-left panel corresponds
to the median of the typical uncertainties in the observed data. Set A refers to the simplest convection model. 
Sets B, C and D additionally include radiative cooling; turbulent pressure and turbulent 
flux; and their combination, respectively.
}
\label{fig:cmd_cep_FU_lmc_th}
\end{figure*}

The CMD for LMC and SMC FU CCs are as shown in Figs.~\ref{fig:cmd_cep_FU_lmc_th} and ~\ref{fig:cmd_cep_FU_smc_th}. In \citet{roma08}, the metallicity range of the CCs span from $-0.18$ dex to $+0.25$ dex for the Galaxy, $-0.62$ dex to $-0.10$ dex for the LMC, and $-0.87$ dex to $-0.63$ dex for the SMC. This variation may further expand with the inclusion of larger spectroscopic data. In the recent study by \citet{bhar24}, the metallicity range for the Galactic sample has already extended from $-0.18$ dex to $-1.1$ dex for metal-poor CCs and from $+0.25$ dex to $+0.6$ dex 
for the metal-rich ones. Hence, we have considered the entire range of chemical compositions from the input grid
while estimating IS edges. The fraction of observed CCs and T2C 
variables (with and without outliers removal) within predicted IS boundaries are shown in Fig.~\ref{fig:star_counts}. It is evident from this figure that 
the IS boundaries predicted from our model grid of CCs are able to include the observed stars within the IS very well. However, more complex convection 
physics captures more FU CCs while these numbers decrease for FO CCs and T2Cs with the increasing complexity of the convection physics. We emphasize that the ``increasing complexity of convection physics'' here refers to the increasing number of turbulent convection parameters integrated into the fiducial convection sets of \textsc{MESA-RSP}. For instance, set A comprises 5 turbulent convection parameters, set B includes 6, set C incorporates 7, and set D comprises 8 parameters. In the calculation of the fraction of observed stars, we have taken into account the uncertainties in intrinsic colour and absolute magnitude. We have propagated these uncertainties into our calculation by averaging over $10^{5}$ random realizations. Specifically, we took into account uncertainties in apparent mean magnitude (derived from the Fourier fit), distance modulus (obtained from 
\citet{piet19,grac20}), and reddening (obtained from \citet{skow21}). We note that the dominant source of uncertainty in absolute magnitude and intrinsic 
colour is coming from the uncertainties in reddening.
% It is evident from this figure that the IS boundaries predicted from our model grid of CCs are able to include the observed stars within the IS very well. 
% For example, it is clear to see from Fig.~\ref{fig:cmd_cep_FU_lmc_th} that $99.95\%$ of OGLE-IV LMC FU CCs are within the 
% set A blue edge and set D red edge.

\begin{figure*}
\includegraphics[width=0.9\textwidth,keepaspectratio]{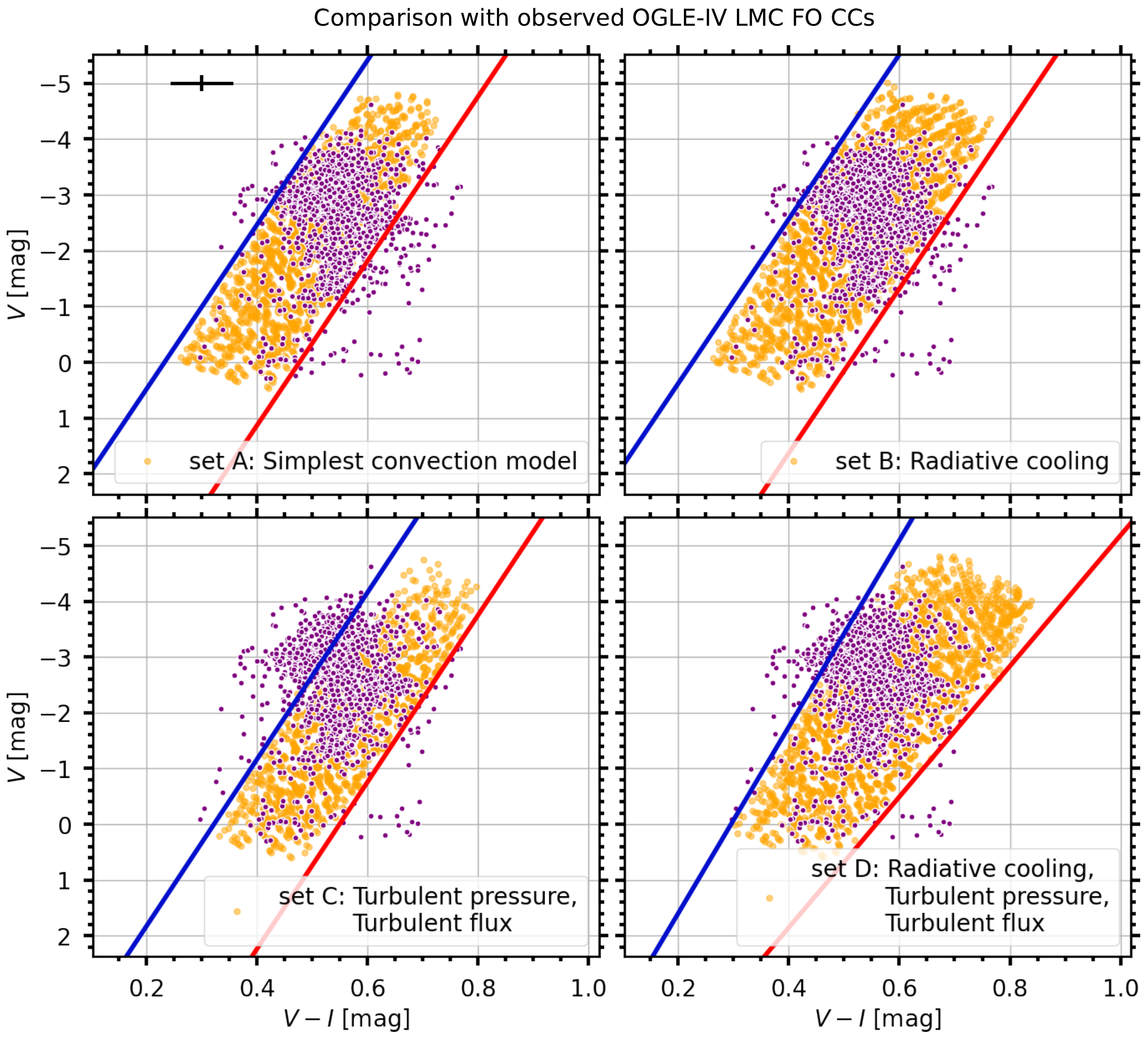}
\caption{Same as Fig.~\ref{fig:cmd_cep_FU_lmc_th}, here with first-overtone Classical Cepheids in the LMC.}
\label{fig:cmd_cep_FO_lmc_th}
\end{figure*}

Similarly, the CMD for LMC and SMC FO CCs are displayed in 
Fig.~\ref{fig:cmd_cep_FO_lmc_th} and Fig.~\ref{fig:cmd_cep_FO_smc_th}, 
respectively. The IS edges exhibit a nearly parallel alignment for FO CCs. 
% This might be due to the higher temperature of the FO CCs, 
% as at higher temperature the effect of turbulent convection parameters gets reduced on the pulsation.

\begin{figure*}
\centering
\includegraphics[width=0.9\textwidth,keepaspectratio]{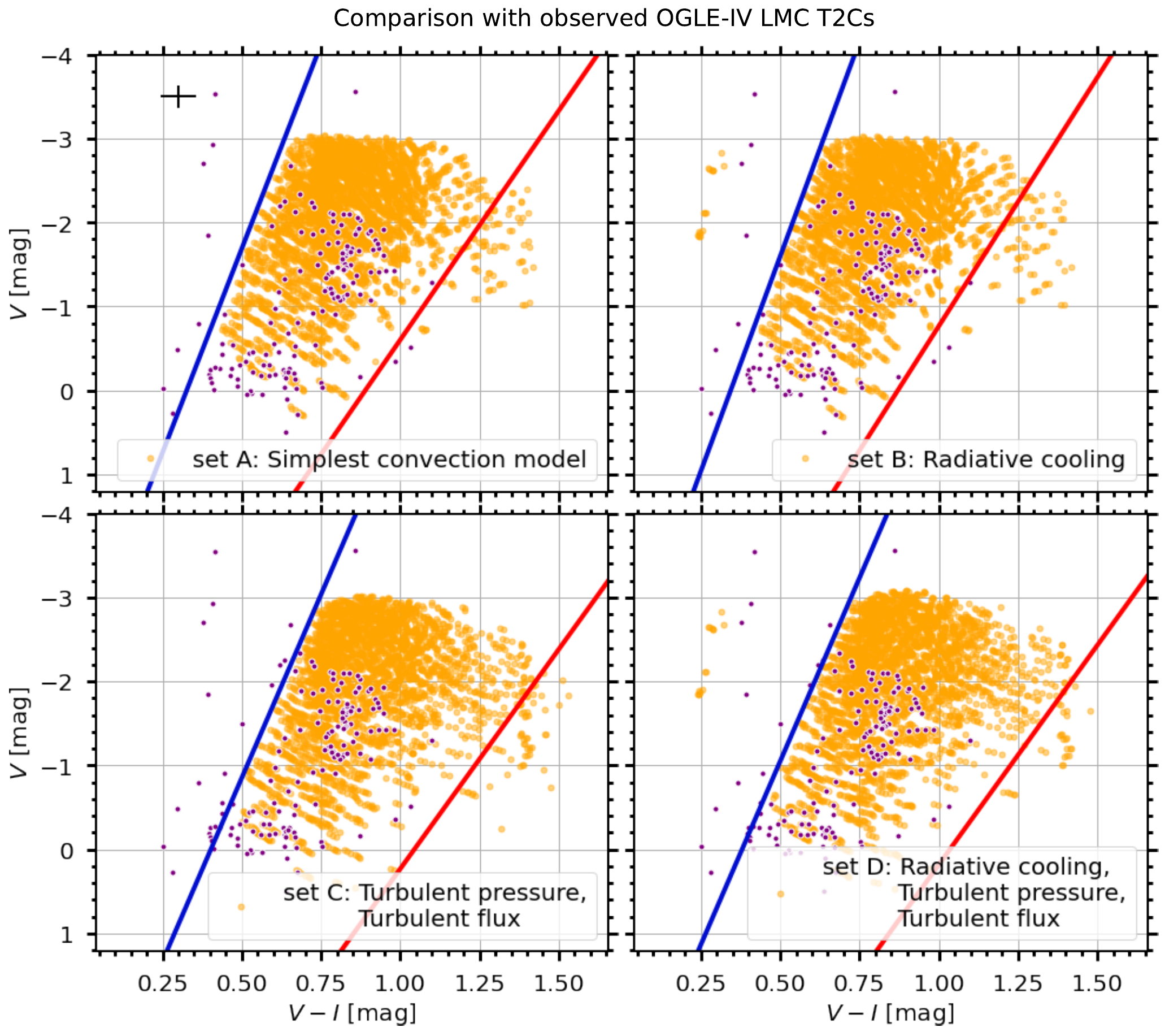}
\caption{Same as Fig.~\ref{fig:cmd_cep_FO_lmc_th}, here with Type-II Cepheids in the LMC.}
\label{fig:cmd_t2cep_lmc_th}
\end{figure*}

\begin{figure*}
\centering
\vspace{0.014\linewidth}
\begin{tabular}{c}
\vspace{+0.01\linewidth}
  \resizebox{1.0\linewidth}{!}{\includegraphics*{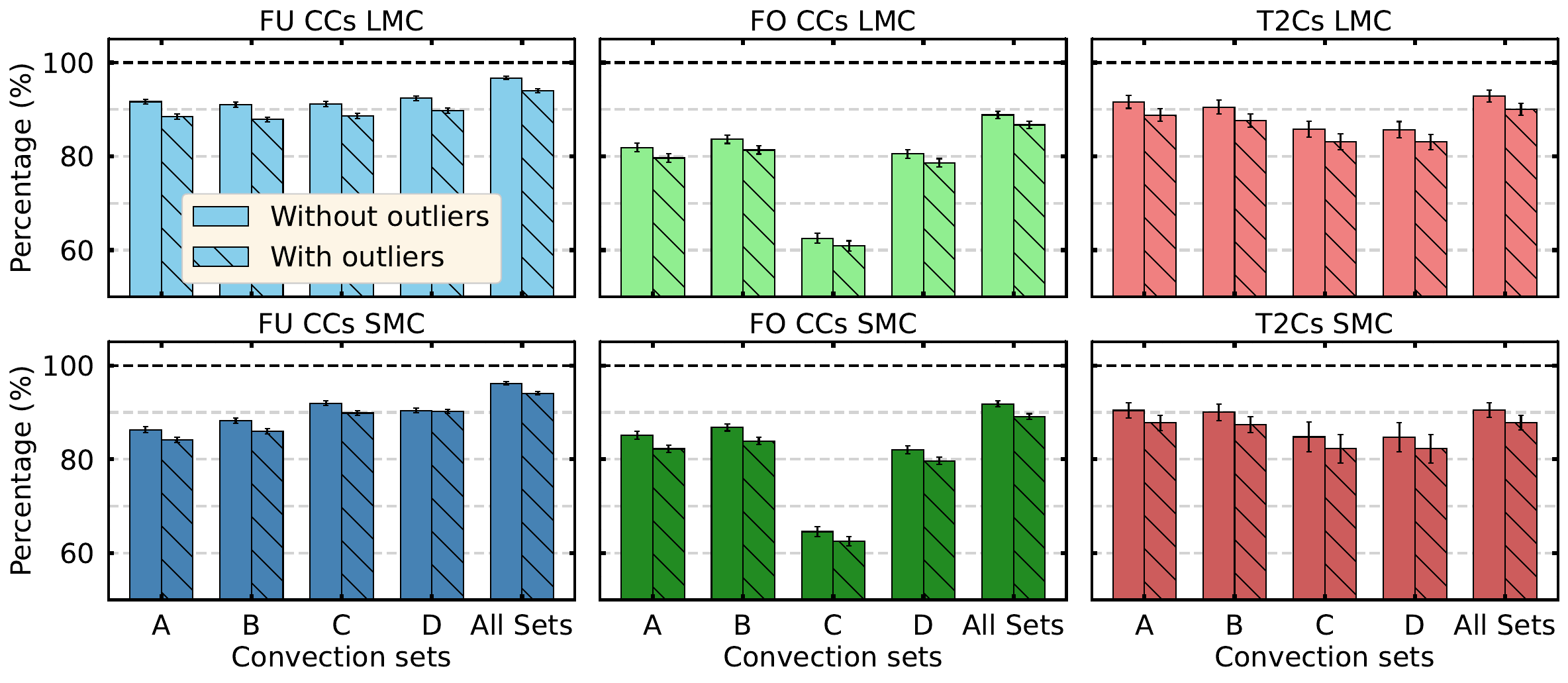}}\\
\vspace{-0.02\linewidth}
\end{tabular}
\caption{Percentage of observed Cepheids within the predicted instability strip across different assumptions in input physics in the convection theory. The upper/lower panels show the LMC/SMC. Fundamental and first-overtone Cepheids are in the left and middle panel, and Type-II Cepheids are in the right panel.}
 \label{fig:star_counts}
\end{figure*}

We further present the comparative CMD for LMC/SMC T2Cs along with their 
predicted edges in Fig.~\ref{fig:cmd_t2cep_lmc_th} and Fig.~\ref{fig:cmd_t2cep_smc_th}, respectively. It is evident
that the edges are becoming redder with the increasing complexity of
convection sets in this case as well.
It should be noted that the number of unstable T2C theoretical models generated in 
the low magnitude regime is also small. This may hint at constraints 
within \textsc{MESA-RSP} or possibly highlight gaps in our understanding and 
selection of turbulent convection parameters.

The slopes and intercepts of the predicted IS edges in colour-magnitude and $L-T_{\rm{eff}}$-plane are listed in Tables~\ref{table:CMD_IS} 
and ~\ref{table:hrd_IS}, respectively. 
It is noteworthy that the proportion of observed stars within each set is fairly consistent, except for the FO CCs set C, which contains nearly $70\%$ of the total observed stars. 
%Fig.~\ref{fig:cep_IS} shows the observed data along with the theoretical IS edges. 

\label{app:FO_CCs}

\begin{center}
 \begin{table*}
        \centering
        \caption{Slopes and intercepts of IS edges for FU CCs, FO CCs and T2Cs in the CMD. These IS boundaries are estimated including the entire range of chemical compositions considered while constructing the two input grids in this work.}
        \label{table:CMD_IS}
        \begin{tabular}{|c|c|c|c|c|c|c|} % four columns, alignment for each
        \hline
		\multirow{4}{*}{Class of Variables}& Convection Sets & \multicolumn{4}{c|}{IS edges  $(V=a(V-I)+b)$}  \\ \cline{3-6}
                           &                 &\multicolumn{2}{c|}{Blue edge} & \multicolumn{2}{c|}{Red edge} \\  \cline{3-6} 
                           &                 & Slope & Intercept & Slope & Intercept \\ \hline
              FU CCs       &A&$-17.568$&$5.493$&$-10.368$&$5.506$ \\ 
			                &B&$-16.522$&$4.941$&$-12.442$&$7.088$ \\
    			            &C&$-17.732$&$6.631$&$-10.048$&$6.426$ \\
    			            &D&$-18.384$&$6.817$&$-9.292$&$5.664$ \\ \hline
	      FO CCs       &A&$-14.752$&$3.433$&$-14.752$&$7.459$ \\
                     &B&$-14.752$&$3.333$&$-14.752$&$7.533$ \\
	                &C&$-14.968$&$4.819$&$-14.968$&$8.219$ \\
			        &D&$-16.724$&$4.943$&$-11.724$&$6.543$ \\ \hline
              T2Cs    &A&$-9.718$&$3.169$&$-5.458$&$4.855$ \\
                      &B&$-10.230$&$3.515$&$-5.926$&$5.159$ \\
	                   &C&$-8.730$&$3.515$&$-5.226$&$5.459$ \\
			             &D&$-8.730$&$3.515$&$-5.226$&$5.409$ \\ \hline
	
        \end{tabular}
\end{table*}

\end{center}

\subsection{Effect of turbulent convection parameters on the IS}
\label{sec:results2}
Here, we discuss the effect of turbulent convection parameters on the predicted IS edges.
We have found that the  IS edges become redder with the increasing 
complexity of the convection sets for both CCs and T2Cs. 
In other words, the set A blue edge is the bluest one and the set D red edge is the reddest one among the IS edges.

Within the set of eight turbulent convection parameters, the values of four parameters ($\alpha,\alpha_{s},\alpha_{c}~ \rm{and} ~\alpha_{d}$) are kept the same for all the four sets. 
The remaining four parameters ($\alpha_{m},\alpha_{p},\alpha_{t}$, and $\alpha_{\gamma}$) assume distinct values for each set.
The effect of set A and set B on the IS edges looks similar, while the same is true for the set C and set D. 
Furthermore, the difference in the topology of IS edges with sets (A and B) and sets (C and D) is very distinct, attributed mainly to the parameters $\alpha_{p}$ and $\alpha_{t}$. 
Turbulent pressure $\alpha_{p}$ contributes to the driving of pulsation, while convective flux $\alpha_{c}$ and eddy viscosity $\alpha_{m}$ act as damping mechanisms. 
We find the inclusion of $\alpha_{p}$ and $\alpha_{t}$ shifts the IS towards a relatively lower temperature as compared to set A and B, reducing the $T_{\rm eff}$ obtained after LNA calculation by $\sim100-200$~K. 
Additionally, they also damp pulsations in some models near the blue edge while driving the same in some models near the red edge. 
Furthermore, \citet{kova23} have found a correlation between $\alpha_{m}$ (in their case, it is denoted as $\alpha_{v}$) and $T_{\rm eff}$, in which
increasing value of $\alpha_{m}$ is suitable for decreasing $T_{\rm eff}$ in case of RR Lyrae stars. In Table~\ref{table:convective_set}, it can be seen that  $\alpha_{m}$ increases from set A to set D except set C. Since we have also found that the IS edges become redder from set A to set D, it may indicate that for CCs also, $\alpha_{m}$ is dependent on $T_{\rm eff}$. Nevertheless, we still do not have a clear picture of the exact contributions of these turbulent convection parameters at low/high temperatures, which we plan to follow up in a future study.
% We have also found that with a decreasing value of effective temperature, a higher value of $\alpha_{m}$ is suitable in agreement with \citet{kova23}.

\begin{figure*}%
\centering%
\includegraphics[width=0.33\linewidth]{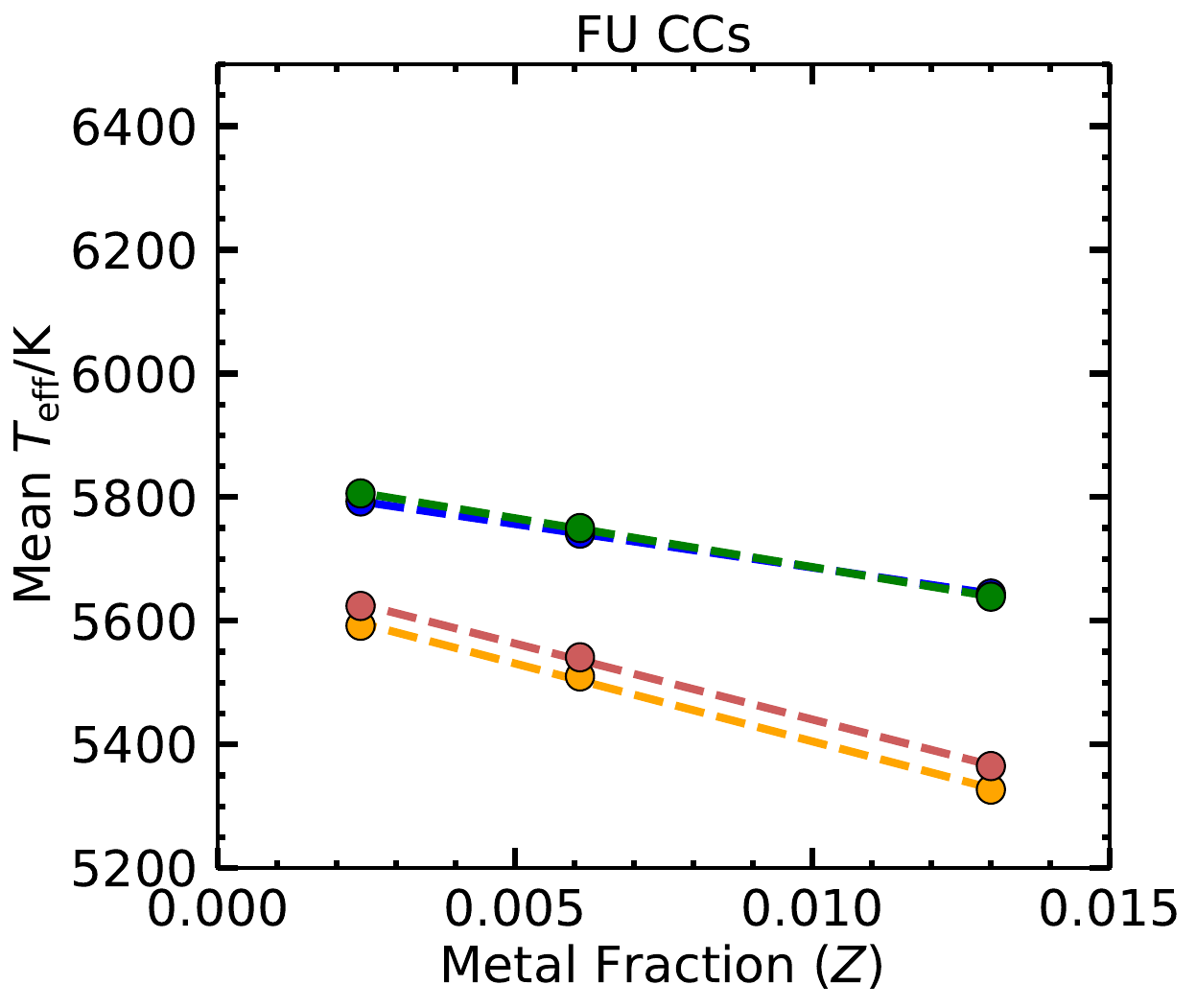}%
\includegraphics[width=0.33\linewidth]{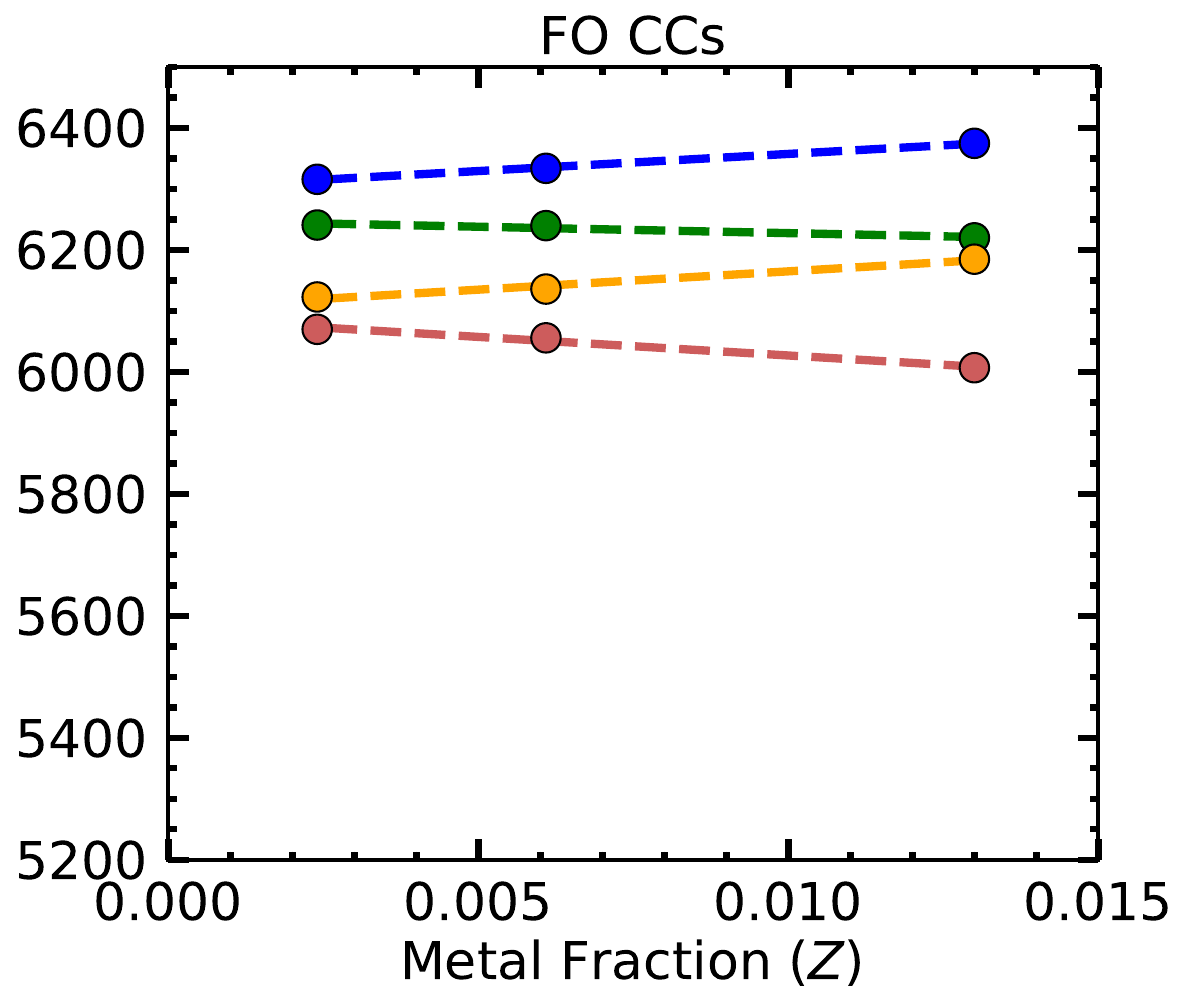}%
\includegraphics[width=0.33\linewidth]{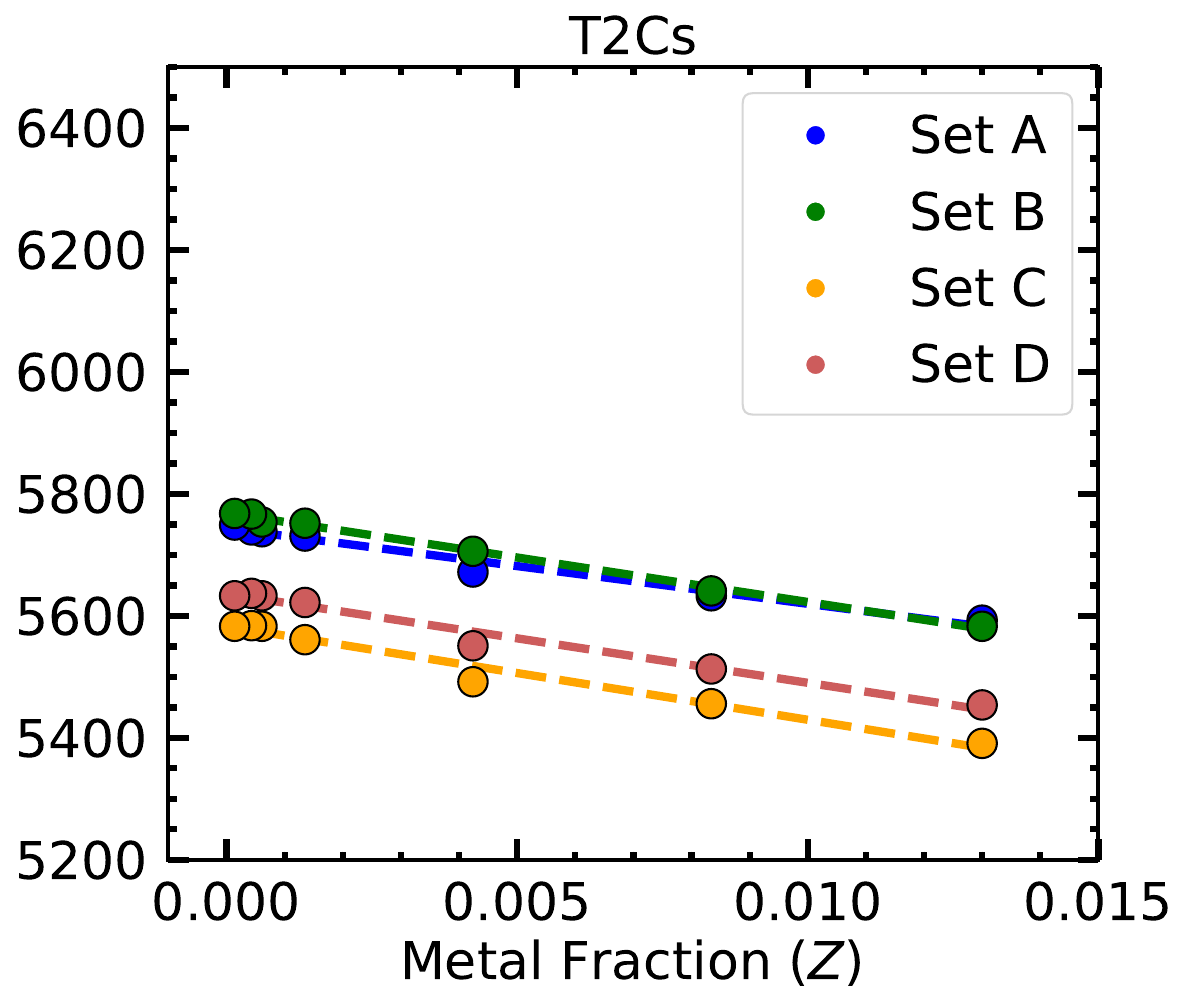}%
\caption{Mean theoretical $T_{\rm eff}$ of Cepheid instability strips, as a function of both metallicity and treatment of turbulent convection. 
The models become cooler with increasing metal fractions as well as with increasing complexity of the convection physics except in a few cases (for example, FO CCs set A and C models).}
\label{fig:quant_temp}
\end{figure*}

To quantify the shift towards lower temperature with increasing complexity of the convection
theory, we have determined the mean effective temperature from each distribution (for each convection set)
by fitting a Gaussian. 
The mean $T_{\rm eff}$ of set A is relatively similar to that 
of set B, while the mean $T_{\rm eff}$ of set C is similar to that 
of set D. The difference in mean $T_{\rm eff}$ from set (A,B) to set (C,D) is $~(150-200)$~K.
The variations in the mean effective temperature of the models across each convection sets and as a function of metal
fractions $Z$ are shown in Fig.~\ref{fig:quant_temp}. 

% \begin{figure}
% \centering
% %\vspace{0.014\linewidth}
% %\begin{tabular}{c}
% %\vspace{+0.01\linewidth}
% %  \resizebox{1.0\linewidth}{!}{\includegraphics*{Figures/t2cep_temp_kde_combined_colored.pdf}}\\
% %\vspace{-0.02\linewidth}
% %\end{tabular}
% \includegraphics[width=\linewidth]{Figures/t2cep_temp_kde_combined_colored.pdf}
% \caption{Effective temperature density distribution of T2Cs with different convection parameter sets. The mean of the distribution 
% shifts towards lower temperatures with increasing complexity of the input physics in the convection theory.}
% \label{fig:quant_temp}
% \end{figure}

\subsection{Comparison with literature}
\label{sec:results3}
\begin{figure*}
\centering
\includegraphics[width=0.35\linewidth]{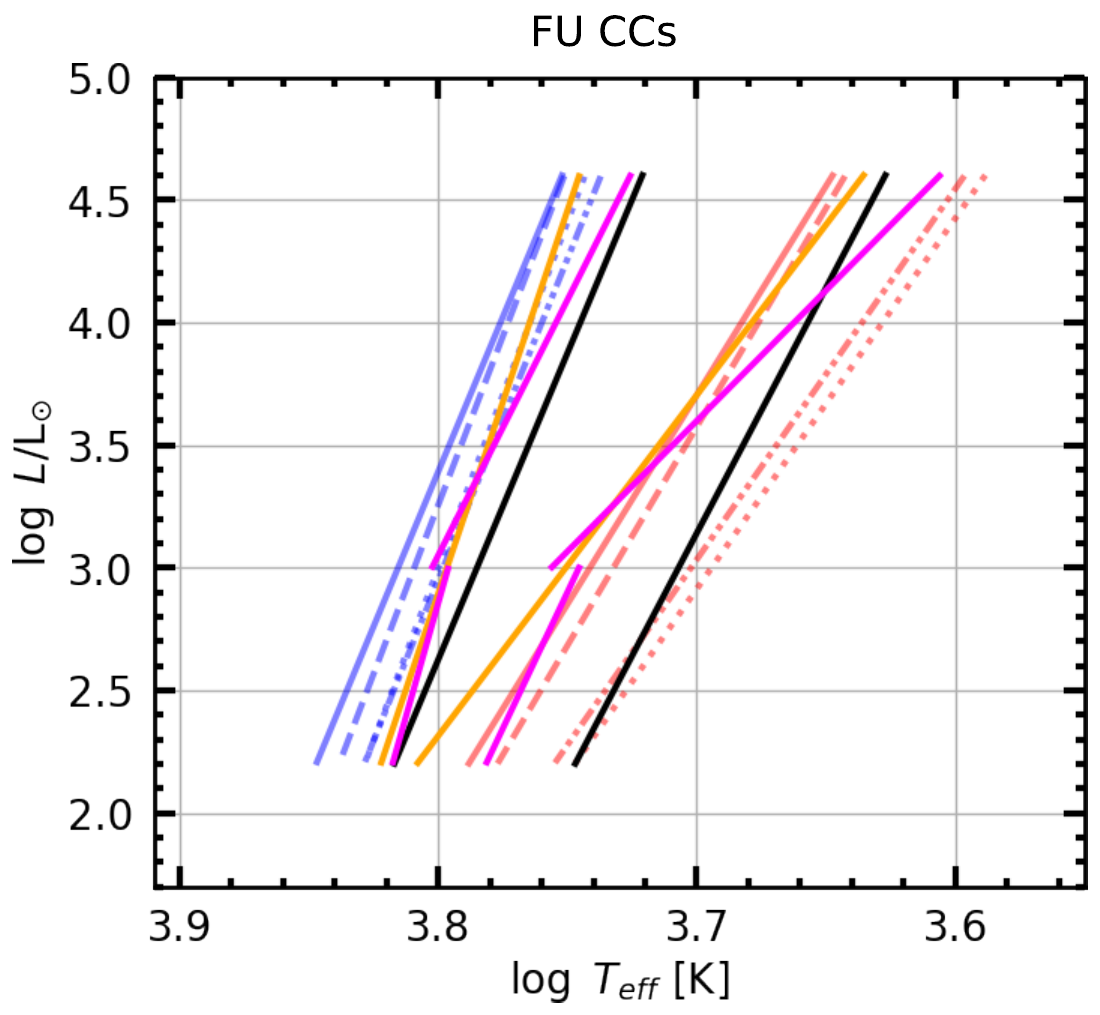}%
\includegraphics[width=0.35\linewidth]{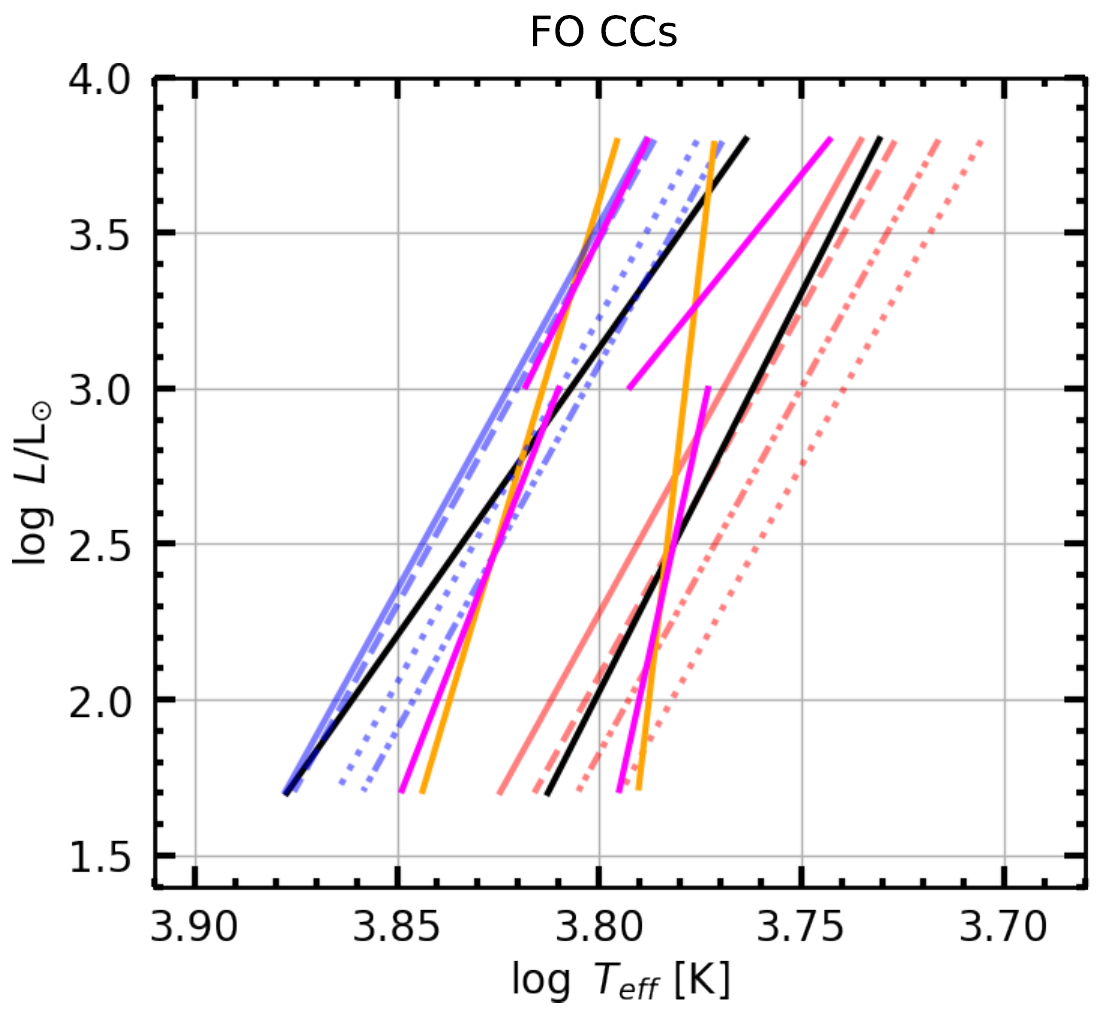}%
\includegraphics[width=0.3\linewidth]{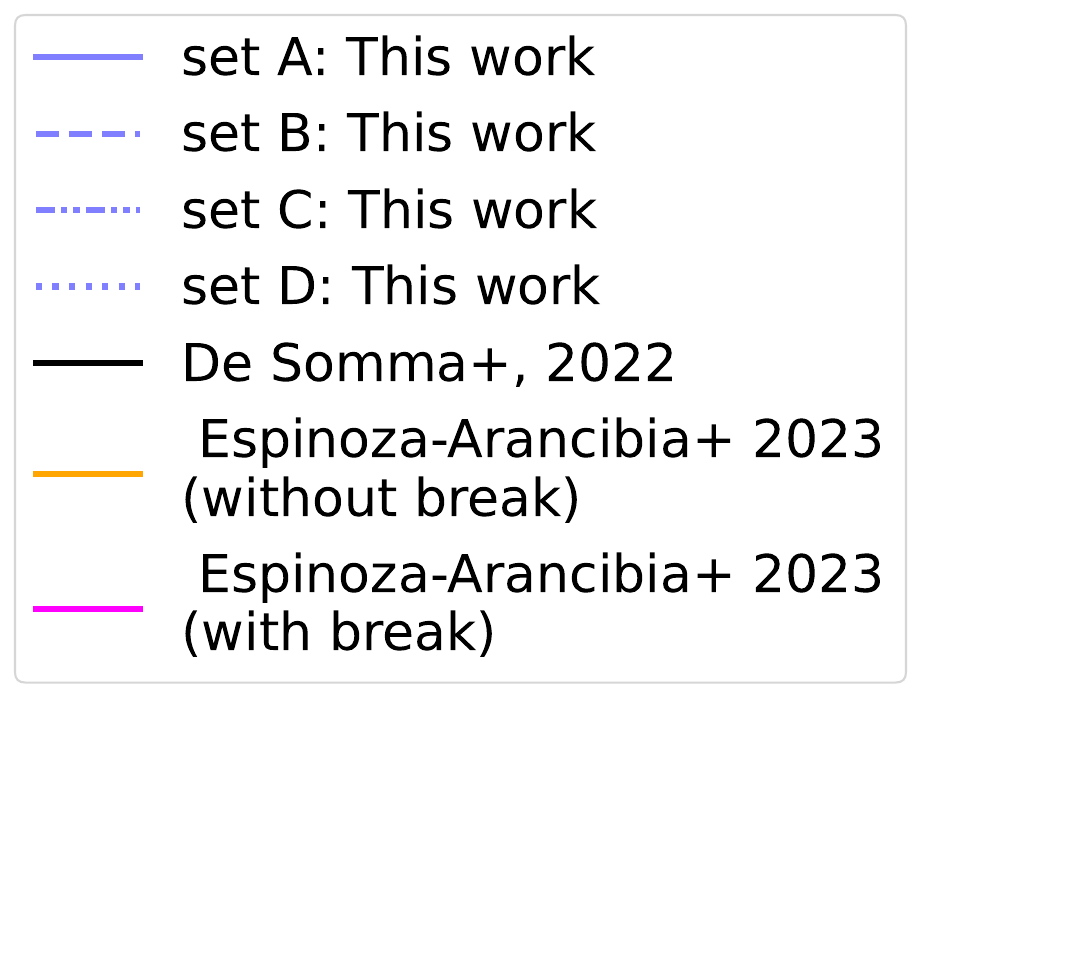}%
\caption{Hertzsprung--Russell diagram of the instability strip (IS) edges for Cepheids across different convection sets (A, B, C, D) from this work compared with recently determined empirical and theoretical IS edges from the literature. 
Red and blue colours represent the theoretical red and blue edges obtained in this work. 
The left panel shows fundamental-mode Classical Cepheids and the right panel shows first-overtone Classical Cepheids. 
The black lines show the theoretical IS edges from \citet{somm22}. 
The orange and magenta lines represent the empirical IS edges from \citet{aran23} with and without a break, respectively.}
\label{fig:hrd_IS}
\end{figure*}

We perform a comparative analysis between our predicted IS edges and those derived from recent investigations by \citet{somm22} and \citet{aran23} in the $LT_{\rm eff}$-plane to better constrain our findings. The study by \citet{somm22} uses non-linear convective theoretical pulsation models, incorporating a mass-luminosity relation based on \citet{bono00a} that excludes the effects of mass loss, rotation, and core-overshooting (referred to as `A'). Additionally, they consider two alternative luminosity levels for the same mass by increasing the luminosity level by 0.2 dex (`B') and 0.4 dex (`C'). The corresponding ISs are computed, accounting for the influences of superadiabatic convection efficiency with $\alpha_{\rm ML}=(1.5,1.7,1.9)$, and varying metallicities $(Z,Y)=(0.004,0.25),(0.008,0.25)$, and $(0.03,0.28)$.
% The IS from \citet{somm22} is over-plotted in this figure for comparison. 
The slopes and intercepts of the IS edges overplotted in Fig.~\ref{fig:hrd_IS} are taken from Table.~10 of
\citet{somm22} with  ML relation `C'  and `B' and $\alpha_{\rm ML}=1.5$ (as referred in their paper) for FU and FO CCs, respectively, as
these IS edges show the closest agreement with our predictions. The slopes show a close match for FU CCs with that of our work, while disparities in intercepts arise from variations in the adopted chemical compositions. Interestingly, for FO CCs, the slope of their red edge aligns more closely with ours, in contrast to their blue edge.
For both FU and FO CCs, the blue edge is observed to be significantly cooler, while the width of the IS appears to be wider as compared to that of \citet{somm22}.

The IS edges computed by \citet{aran23} are derived from the same observed data employed in our study. They have determined the IS edges both without breaks and with breaks introduced at a period of $\sim 3$ days. 
For FU CCs, both the upper and lower empirical blue edges by \citet{aran23} align well with our set D blue edge, whereas the lower empirical red edge corresponds closely to our set A red edge. Notably, the upper empirical red edge from \citet{aran23} falls between our set A and set D red edges. 

In the case of FO CCs, the empirical upper blue edge aligns well with our
estimated set A/B blue edges, while the lower empirical blue edge corresponds 
closely to our set C/D blue edges. However, we can see some distinct differences 
in our predicated red edges with their empirical lower/upper red edges. 
This also highlights the importance of optimising the current sets of turbulent convection parameters in \textsc{MESA-RSP} as the position of the red edge is highly dependent on the complex interaction between convection and pulsation.

\subsection{Effect of metallicity}
\label{sec:results4}
We have also investigated the effect of metallicity on the IS boundaries for all convection sets. The slopes and intercepts of IS edges
corresponding to different metallicities are listed in Table~\ref{table:metal_IS}. It has been observed that the IS edges generally exhibit a redder shift with increasing 
metallicities for both CCs and T2Cs, with a few exceptions in specific convection set scenarios. This is found to be consistent with \citet{somm22}. 
This is a consequence of the enhanced opacity in stellar envelopes with higher metal content.
%, which subsequently results in a more pronounced efficiency in damping 
%pulsations. 
Specifically, an increase in metal fraction ($Z$) shifts the blue edge towards lower temperature, while 
the increase in helium fraction ($Y$) shifts the blue edge towards higher temperature \citep{cox80}. In our model 
grids, the $Y$ values increase with an increase in the $Z$ values.  Therefore, we observe a 
very subtle effect of metallicity on the computed IS edges. The mean effective temperature ($T_{\rm eff}$) is quantified 
for the model distribution for the individual metallicity values and is shown in Fig.~\ref{fig:quant_temp}. 
% \begin{figure*}
% \vspace{0.014\linewidth}
% \begin{tabular}{c}
% \vspace{+0.01\linewidth}
%   \resizebox{1.0\linewidth}{!}{\includegraphics*{Figures/metal_depend_IS.pdf}}\\
% \vspace{-0.02\linewidth}
% \end{tabular}
% \caption{Effect of metal fraction ($Z$) on the instability strip edges for FU and FO mode Classical Cepheids (CCs) and Type-II Cepheids (T2Cs) across different convection sets. Blue and red colours represent the blue and red edges, respectively.
% %Although for the T2Cs theoretical grid, seven different metallicity values are considered,  we have shown the IS edges for only three metallicity values (highest, middle and lowest one) in the plot for visualization purposes.
% }
% \label{fig:metalcep_IS}
% \end{figure*}

\subsection{Flare in the theoretical colour-magnitude diagram}
\label{sec:results5}
\begin{figure}
\vspace{0.014\linewidth}
\begin{tabular}{c}
\vspace{+0.01\linewidth}
  \resizebox{0.9\linewidth}{!}{\includegraphics*{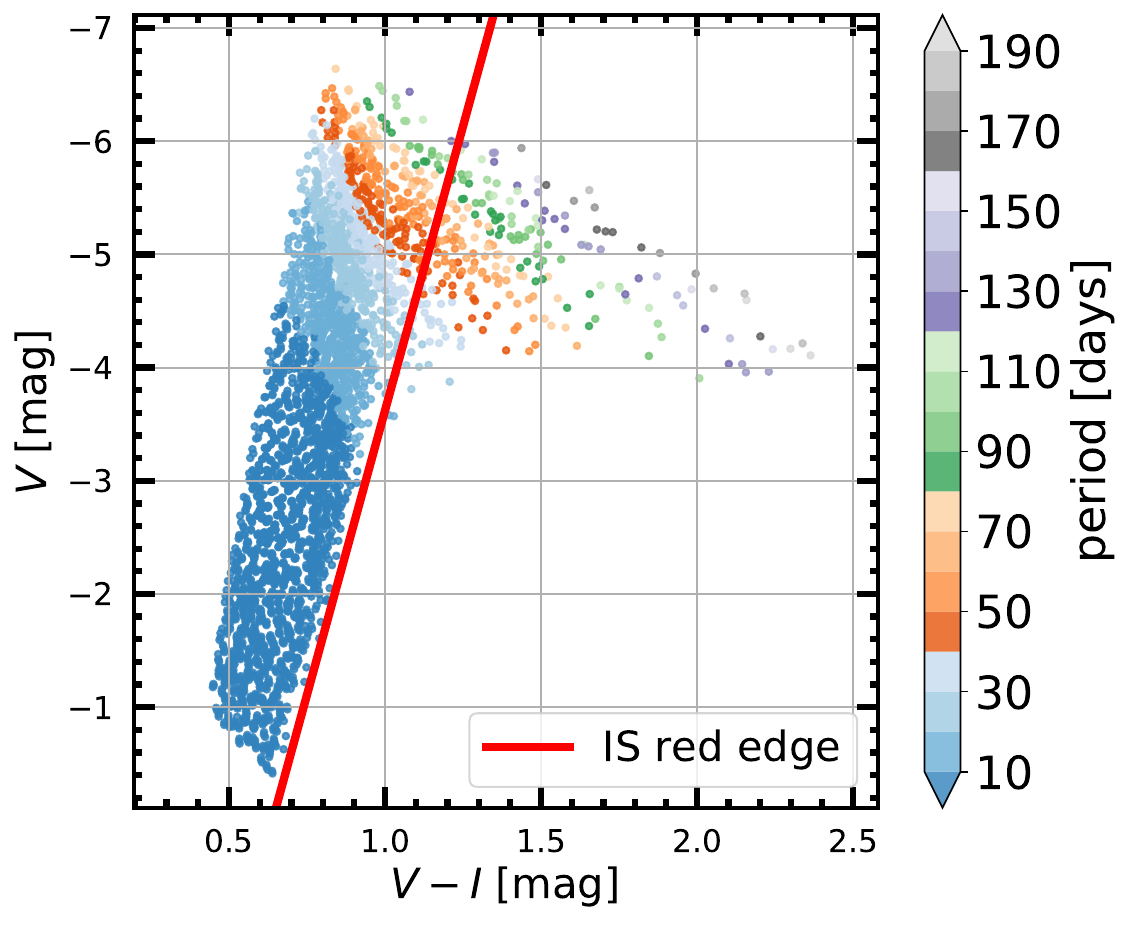}}\\
\vspace{-0.02\linewidth}
\end{tabular}
\caption{The CMD for theoretical models with set C. The red line represents the red edge predicted for this particular set. 
Unstable models with ${\rm period>20}$~days (which we refer to as flared models) can be seen in the low $T_{\rm eff}$ and high $L$ region beyond our predicted IS red edge.}
\label{fig:flare}
\end{figure}

Unstable models show ``flaring'' in the low $T_{\rm eff}$ and high $L$ region beyond the predicted IS red edge in the present work, as shown in Fig.~\ref{fig:flare}. 
These flares are seen in models with periods greater than $20$~ days with all four convection sets, A-D. 
The flare is relatively larger in sets C and D as compared to sets A and B. 
However, the observed data do not show such flares. 
The effects of convection become very significant near the red edge of IS \citep{smol08}.  

Some preliminary findings from the flared models: 
\begin{enumerate}
    \item The flared models consist of $\sim6\%$ of the total pulsating models generated in this work.
    \item The coolest of the flared models extend nearly to the Hayashi limit.
    \item With increasing metallicity, the number of flared models tends to increase.
    \item Inclusion of radiative cooling $\gamma_{r}$ reduces the number of flared models.
    \item The pulsation can be damped in the flared models by increasing the value of eddy viscosity $\alpha_{m}$ and convective flux $\alpha_{c}$ from the fiducial values. 
    Notably, adjusting $\alpha_{m}$ has minimal impact on the linear pulsation period, while changes in $\alpha_{c}$ significantly influence the period. 
    This also suggests the necessity of establishing $\alpha_{m}$ as a function of $T_{\rm eff}$ for CCs similar to RRLs \citep{kova23,kova24}.
\end{enumerate}

\subsection{Cepheid evolutionary models}
\label{sec:results6}
\begin{figure}
\centering
\includegraphics[width=\linewidth,keepaspectratio]{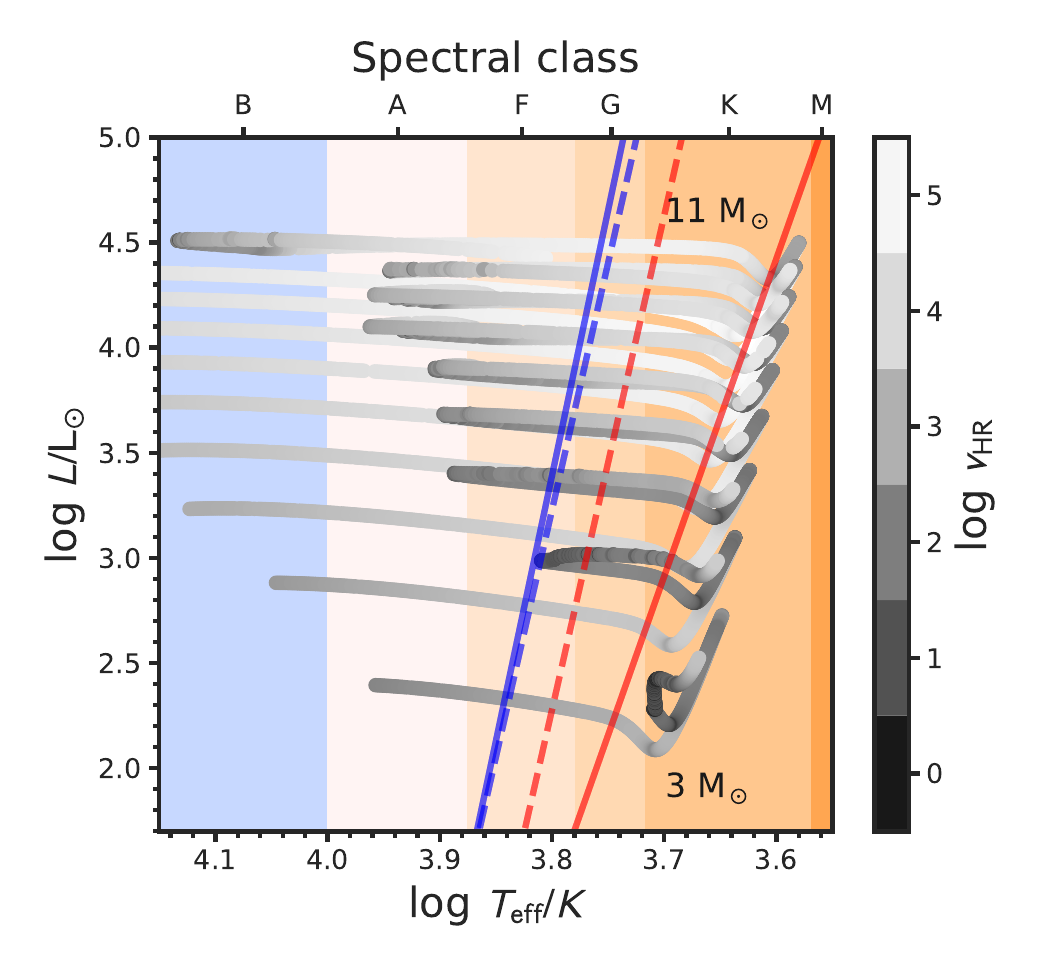}
\caption{Cepheid evolutionary tracks are shown from core~H exhaustion to core~He exhaustion.
The tracks are computed with masses $(3-11)~\rm M_{\odot}$ in steps of $1~\rm M_{\odot}$. 
The FU (solid lines) and FO (dashed lines) IS edges as given in Section~\ref{sec:summary} are overplotted,
where blue and red colours represent blue and red edges respectively. 
}
\label{fig:ev_track}
\end{figure}

To show the consistency of the predicted IS edges with the evolutionary models, we have computed stellar evolutionary tracks for Cepheids with mass range $(3-11)~\rm M_{\odot}$ in steps of $1~\rm M_{\odot}$ using \textsc{MESA}~r15140. 
We have adopted the LMC composition for the calculation (see Section~\ref{sec:th_data} and Table~\ref{table:feh}). We treat convection using the mixing-length theory (MLT) from \citet{heny65} and the boundary of the convective region is determined using the Ledoux criterion. We use the step overshooting prescription integrated in MESA. 
We use a value of $0.2$ pressure scale heights for incorporating overshooting of core and envelope convection zones. We have implemented the overshooting during the core hydrogen burning and shell non-burning stages.
To include the effects of semi-convection \citep{lang83}, we use an efficiency parameter of $0.1$ for models with mass less than $9~\rm M_{\odot}$ and $100$ for masses $9-11~\rm M_{\odot}$.
The evolution is computed from the pre-main sequence stage till the completion of 
the blue loop. The computed tracks are shown in Fig.~\ref{fig:ev_track} along with the 
predicted IS edges for CCs. For this computation, we have used the \emph{5M\_cepheid\_blue\_loop} inlist from the \textsc{mesa}-r15140 \emph{test\_suite} with a few modifications. For details, see the inlists available at \url{https://github.com/mami-deka/LNA_CMD}.
We note that the evolutionary tracks computed for this project are indicative of the evolutionary path followed by CCs. The calibration of $\alpha_{\rm MLT}$ and overshooting parameters with observation are beyond the scope of the present study.

Additionally, to explain the dearth of long-period observed FU CCs and the flared theoretical models, we determine the speed of the evolutionary models in the HRD using the formulation as given in \citet{bell23}:
\begin{align}
\nu_{\rm HR}=\frac{1}{C\Delta t} \sqrt{[\Delta \log(L/\rm L_{\odot})]^{2} + [\Delta \log(T_{\rm eff}/K)]^{2}}.
\end{align}
Here, we set the normalization factor $C$ to be the zero-age main sequence (ZAMS) velocity of a $3~\rm M_{\odot}$ star. 
The ``speed'' here is not a physical velocity but rather indicates how quickly the star moves through the HRD throughout its evolution, and hence indicates the relative longevity of stellar models in different phases of evolution.
For instance, a $\log{\nu_{\rm HR}}$ value of 0 within a particular region of the HRD implies an expectation of the same stellar population density as that of a $3 \rm M_{\sun}$ star at the ZAMS, assuming the birthrate is the same. Conversely, a $\log {\nu_{\rm HR}}$ value of 3 indicates an expectation of a thousand times fewer stars in that region \citep{bell23}. 
% Therefore, a higher $\log {\nu_{\rm HR}}$ value for a specific model suggests a lower likelihood of that model occupying that position in the HRD \citep{bell23}. 
We can thus explain the existence of CCs in the blue loops from Fig.~\ref{fig:ev_track} as the models in this region have the lowest speeds. However, for high-mass (long-period) CCs, the speed of the models increases relatively even during the blue-loop phase, which accounts for the fewer observed CCs in that region in addition to the observed initial mass function which has few massive stars. 
This could also offer a plausible explanation for the absence of observed stars in the flared region.

\section{Summary and Conclusions}
\label{sec:summary}
We have carried out a comparative study between the observed and theoretical CMDs of FU CCs, FO CCs and T2Cs. We used the observed data from the OGLE-IV database and generated the theoretical data using \textsc{MESA-RSP}. 
We constructed dense theoretical grids of 
pulsation models with stellar parameters $M, L, T_{\rm eff}, Z, X$ and different sets of turbulent convection parameters for both CCs and T2Cs. 
The grids are constructed in such a way that they produce approximately similar values for the minimum and maximum magnitude/colour limits to those of the observed target stars. 
Then, using an ML algorithm, we estimated the IS edges from the theoretical data. 
The results obtained from the present analysis are summarized as follows:
\begin{enumerate}
\item We find that $\sim 90\%$ of observed stars (CCs and T2Cs) fall within the predicted Set A blue edge and Set D red edge. 
We find the following theoretical IS boundaries to include the majority of the observed stars:
\begin{align}
\mathrm{FU~ CCs~ blue~ edge (from~set~A)}:V=&-17.568(V-I)+5.493 \\
\mathrm{red~ edge (from~set~D)}:V=&-9.292(V-I)+ 5.664\\
\mathrm{FO~ CCs~ blue~ edge (from~set~A)}:V=&-14.752(V-I)+3.433 \\
\mathrm{red~ edge (from~set~D)}:V=&-11.724(V-I)+8.652\\ 
\mathrm{T2Cs~ blue~ edge (from~set~A)}:V=&-9.718(V-I)+ 3.169 \\
\mathrm{red~ edge (from~set~D)}:V=&-5.226(V-I)+ 5.409.   
\end{align}
\item We find discrepancies between theoretical and observed CMDs of FU CCs in the low-effective 
temperature and high luminosity regions for stars with periods greater than $\sim 20$ days. 
We find unstable theoretical models in that region where no observed stars are found yet. 
These flared models might be useful for optimizing the turbulent convection parameters for CCs, or for potentially predicting a class of rare, long-period, red Cepheids. 
However, these predictions may only be a consequence of the chosen turbulent convection parameters, and may not be realized in nature. 
It is worth noting that if these red Cepheids do exist, then these results suggest they may arise through an evolutionary path that is distinct from that of CCs, namely, outside of blue loops while high up on the red giant branch. They may represent a new class of variable stars, or 
potentially belong to a previously recognized class of pulsating stars, such as the one found by \citet{clar15}. A detailed investigation will provide a deeper insight into their nature which we intend to carry out in a follow-up study. However, regardless of their classification, they are expected to exhibit a PL relation that is noticeably less steep than that of FU CCs (see Fig.~\ref{fig:PL_cep}).
%\textbf{However, the red Cepheids might be a consequence of the chosen turbulent convection parameters, and they may not be realized in nature. Another possibility is that these stars might have become Cepheids outside of blue loops, even while on the red giant branch. Then, they might quickly enter the blue loop phase where they have a longer life-span, raising questions about their existence on the red giant branch.}
\item The IS edges tend to become redder as the complexity of the convection sets increases. The locations 
of the IS edges (more specifically the red edge) are highly influenced by the turbulent pressure and turbulent flux parameters.
The present analysis also indicates a correlation between $\alpha_{m}$ and $T_{\rm eff}$.
Hence, this indicates that the appropriate turbulent convection parameters may vary as a function of stellar parameters. 
\item The IS edges also become redder with increasing metallicities, except in a few cases involving specific convection sets.
\item \textsc{MESA-RSP} is not able to generate T2C theoretical pulsating models similar to the observations in lower magnitude-lower 
temperature region with the fiducial turbulent convection and numerical parameters. 
Detailed investigation and calibration of the parameters involved are required in this regard. 
\end{enumerate}

The present work is based on a linear analysis of CC and T2C models obtained using \textsc{MESA-RSP}.  
We have found that the present two grids of models are able to reproduce the average properties of the observed stars very well. 
Additionally, it is evident that the combination of different sets of turbulent convection parameters is necessary to reproduce the observed data. 
%Furthermore, this study opens up the possibility of optimizing the turbulent convection parameters using the flared models. 
We plan to investigate the effect of turbulent convection parameters on the light curves using non-linear computation in a future work. 
We also plan to extend this analysis from the optical to the near-infrared band in the future in order to better constrain the physics of these stars.

\section*{Acknowledgements}
The authors thank the anonymous referee for the valuable comments and suggestions which have significantly improved the manuscript.
This work is partially funded by CSIR through a research grant ``03(1425)/18/EMR-II''. 
The authors acknowledge the use of High-Performance Computing facility Pegasus at IUCAA, Pune. 
This study made use of \textsc{MESA}~r15140 
\citep{paxt10,paxt13,paxt15,paxt18,paxt19}. 
MD thanks Gautam Bhuyan and Felix Ahlborn for useful discussions.
S. Das acknowledges the KKP-137523 `SeismoLab' \'Elvonal grant of the Hungarian Research, Development and Innovation Office (NKFIH).

%%%%%%%%%%%%%%%%%%%%%%%%%%%%%%%%%%%%%%%%%%%%%%%%%%
\section*{Data Availability}
The observed data underlying this article are available at \url{http://ftp.astrouw.edu.pl/ogle/ogle4/OCVS/lmc/cep/},
\url{http://ftp.astrouw.edu.pl/ogle/ogle4/OCVS/smc/cep/} and \url{http://ogle.astrouw.edu.pl/}. Theoretical models were computed using MESA r15140. The \textsc{MESA-RSP} inlists used can be found at \url{https://github.com/mami-deka/LNA_CMD}. The derived data generated in this research will be shared on reasonable request to the corresponding author.



\bibliographystyle{mnras}
\bibliography{mnras} % if your bibtex file is called example.bib

% %%%%%%%%%%%%%%%%%%%%%%%%%%%%%%%%%%%%%%%%%%%%%%%%%%
%\newpage
% %%%%%%%%%%%%%%%%% APPENDICES %%%%%%%%%%%%%%%%%%%%%
\appendix
\section{Additional figures}
\begin{figure*}%[HT]
\vspace{0.014\linewidth}
\begin{tabular}{cc}
\vspace{+0.01\linewidth}
  \resizebox{0.45\linewidth}{!}{\includegraphics*{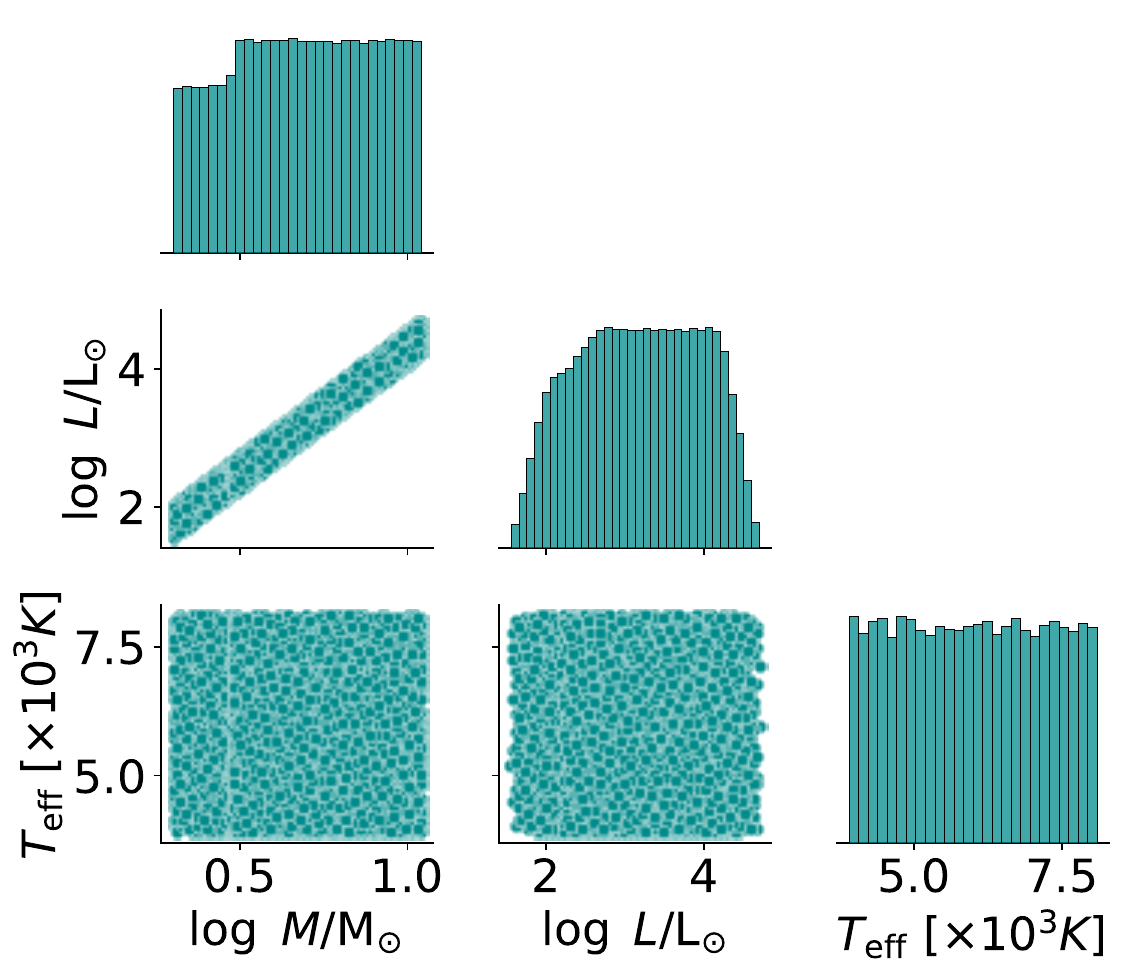}}&
  \resizebox{0.45\linewidth}{!}{\includegraphics*{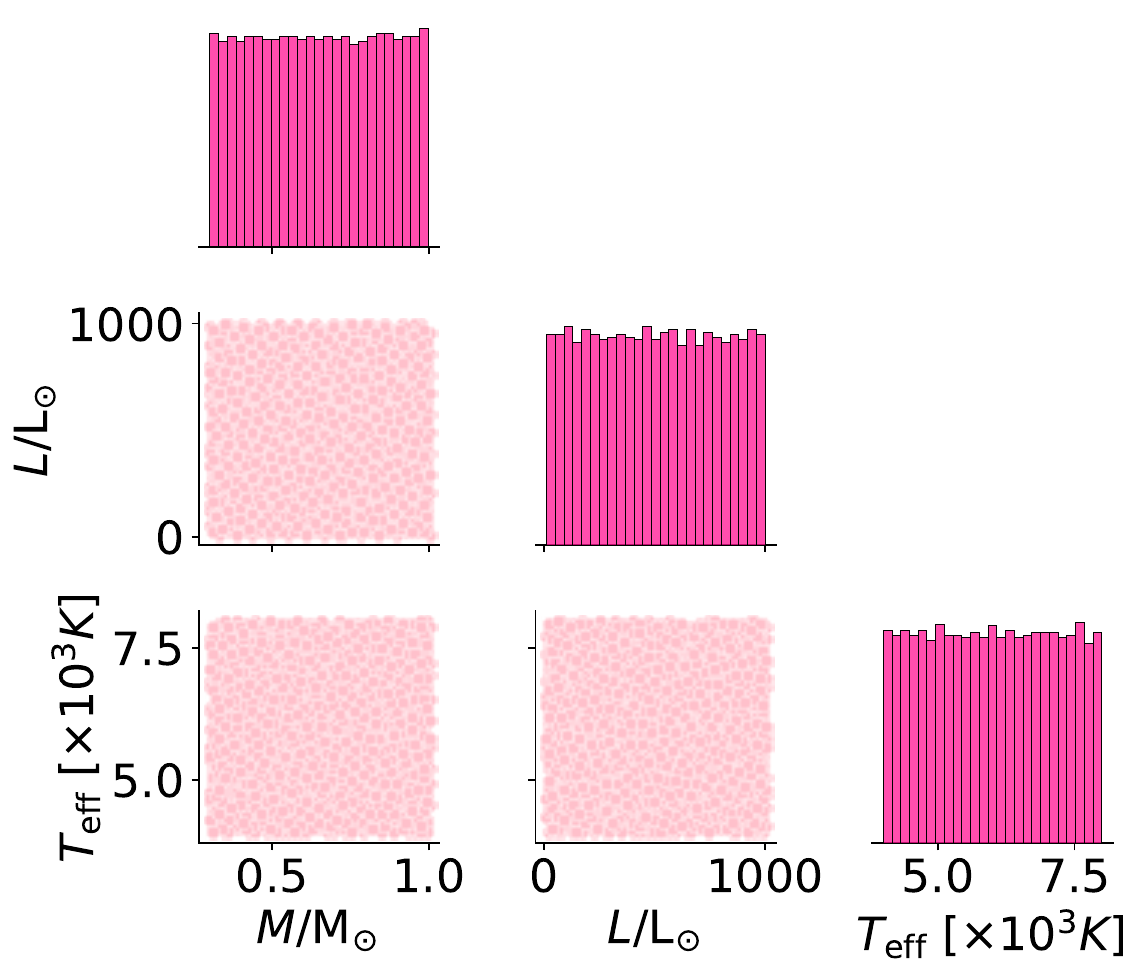}}\\
\vspace{-0.02\linewidth}
\end{tabular}
\caption{Scatterplot matrix of the input stellar parameters: mass ($M$), luminosity ($L$) and temperatures ($T_{\rm eff}$) for CCs (left panel) and T2Cs (right panel). The observed underdensity of low-mass stars in the CC grid (left panel) can be attributed to the grid's construction in two segments. Specifically, the grid was divided into two parts: the first part encompassing stellar masses ranging from $(3-11)\rm M_{\odot}$ and the second part covering $(2-3)\rm M_{\odot}$. %The extension of the original grid with the second part was necessary to cover the observed stars in the lower magnitude region of the CMD. 
}
\label{fig:input_grid}
\end{figure*}

\begin{figure*}
\includegraphics[width=0.9\textwidth,keepaspectratio]{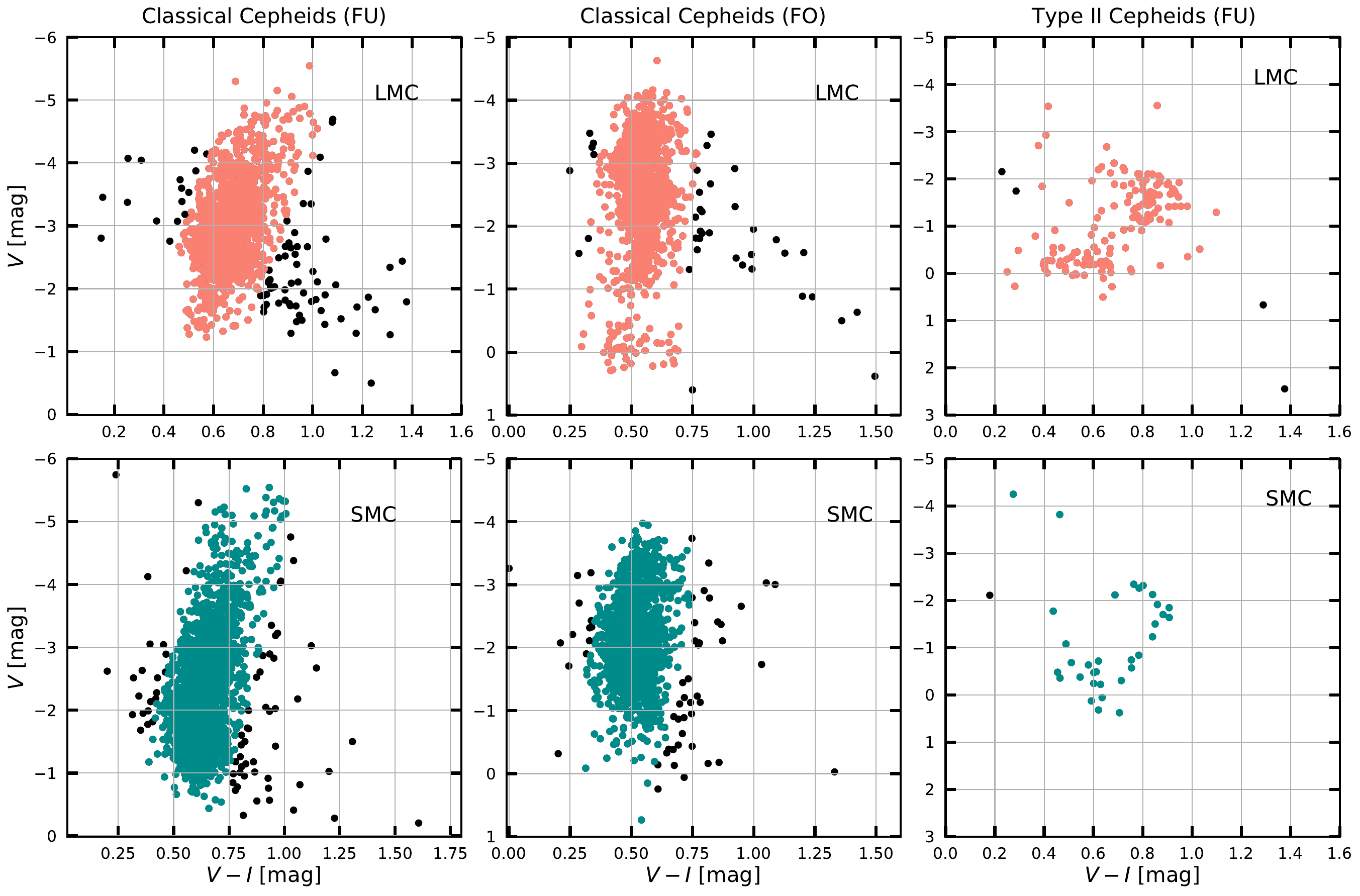}
\caption{CMDs for FU CC, FO CC and T2C stars. 
Black points represent the stars rejected following \citet{mado17} due to their high horizontal deviation in the CMD. 
% Uncertainties in reddening and colour estimation may be responsible for these deviations.
}
\label{fig:obs_cmd}
\end{figure*}

Fig.~\ref{fig:input_grid} shows the scatterplot matrix for the two input stellar parameters grids constructed for the present study and the
observed CMDs are shown in Fig.~\ref{fig:obs_cmd}. 

\section{Sensitivity of the numerical parameters}
\begin{figure*}
\vspace{0.014\linewidth}
\begin{tabular}{c}
\vspace{+0.01\linewidth}
  \resizebox{0.9\linewidth}{!}{\includegraphics*{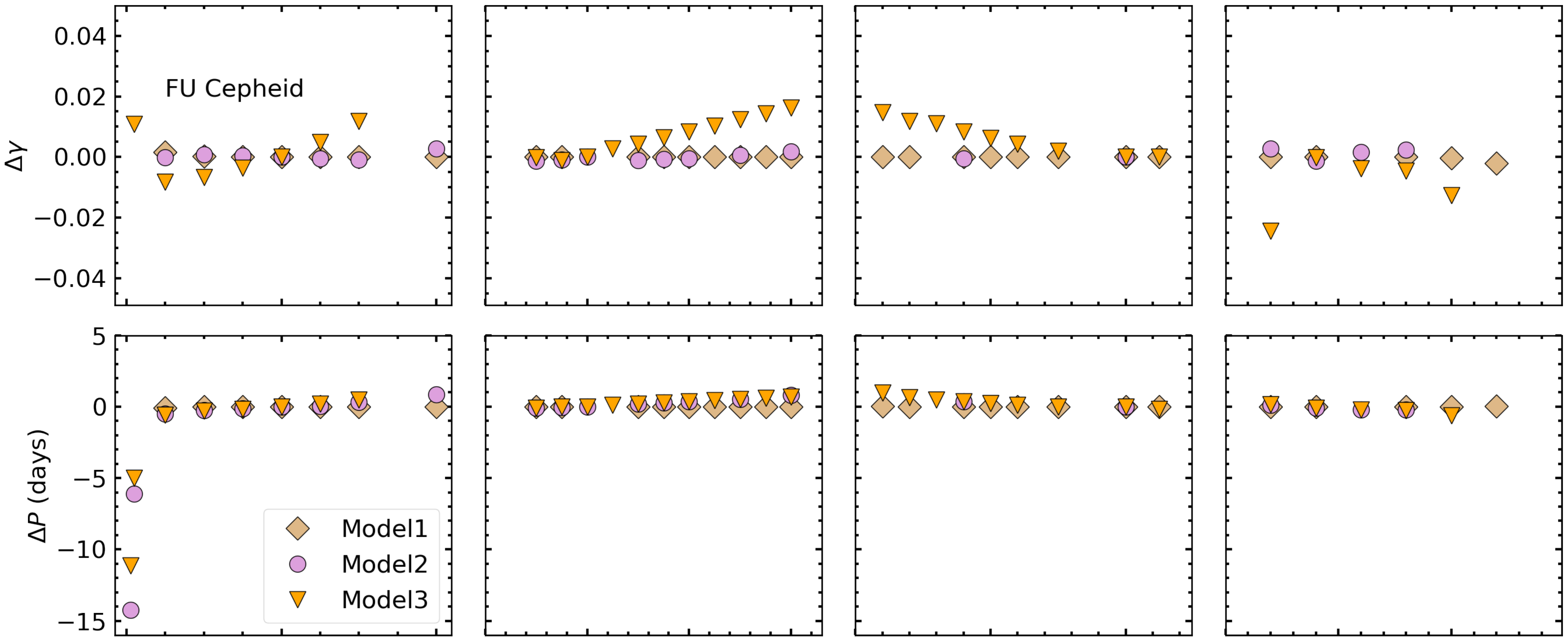}}\\
  \resizebox{0.9\linewidth}{!}{\includegraphics*{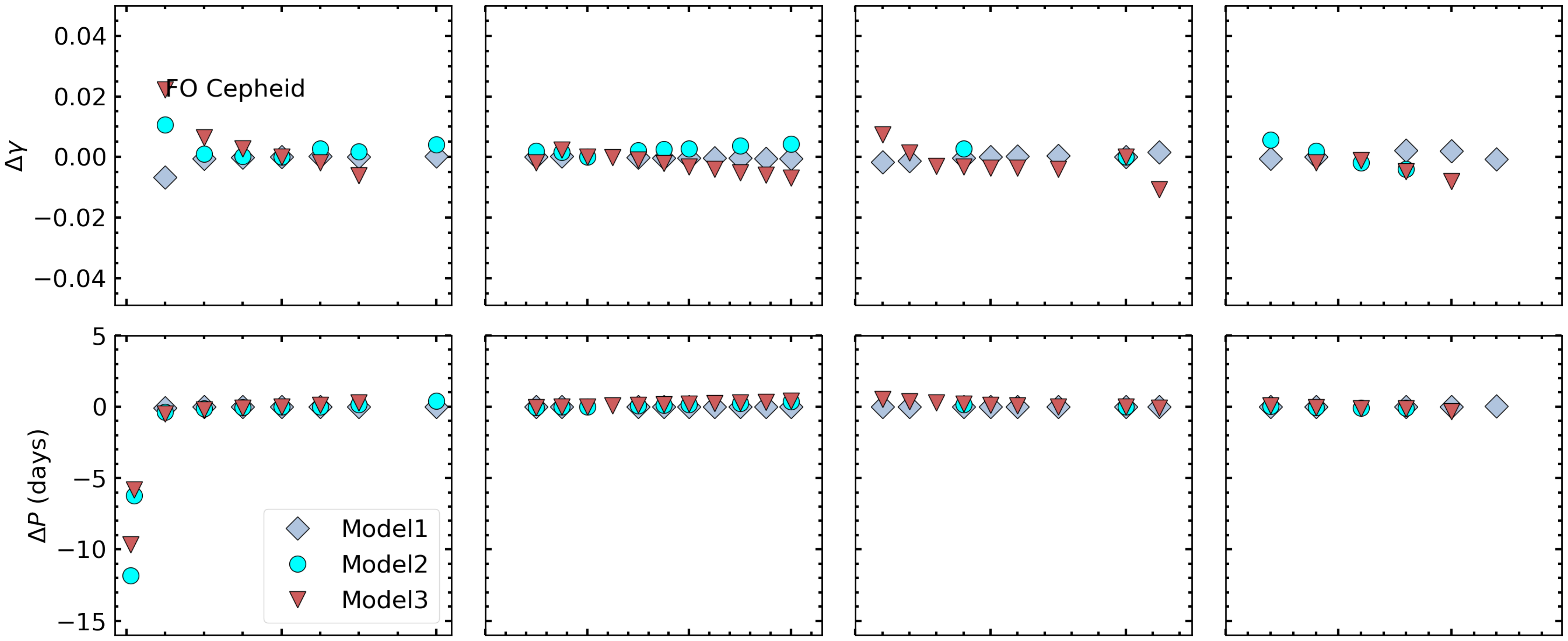}}\\
  \resizebox{0.9\linewidth}{!}{\includegraphics*{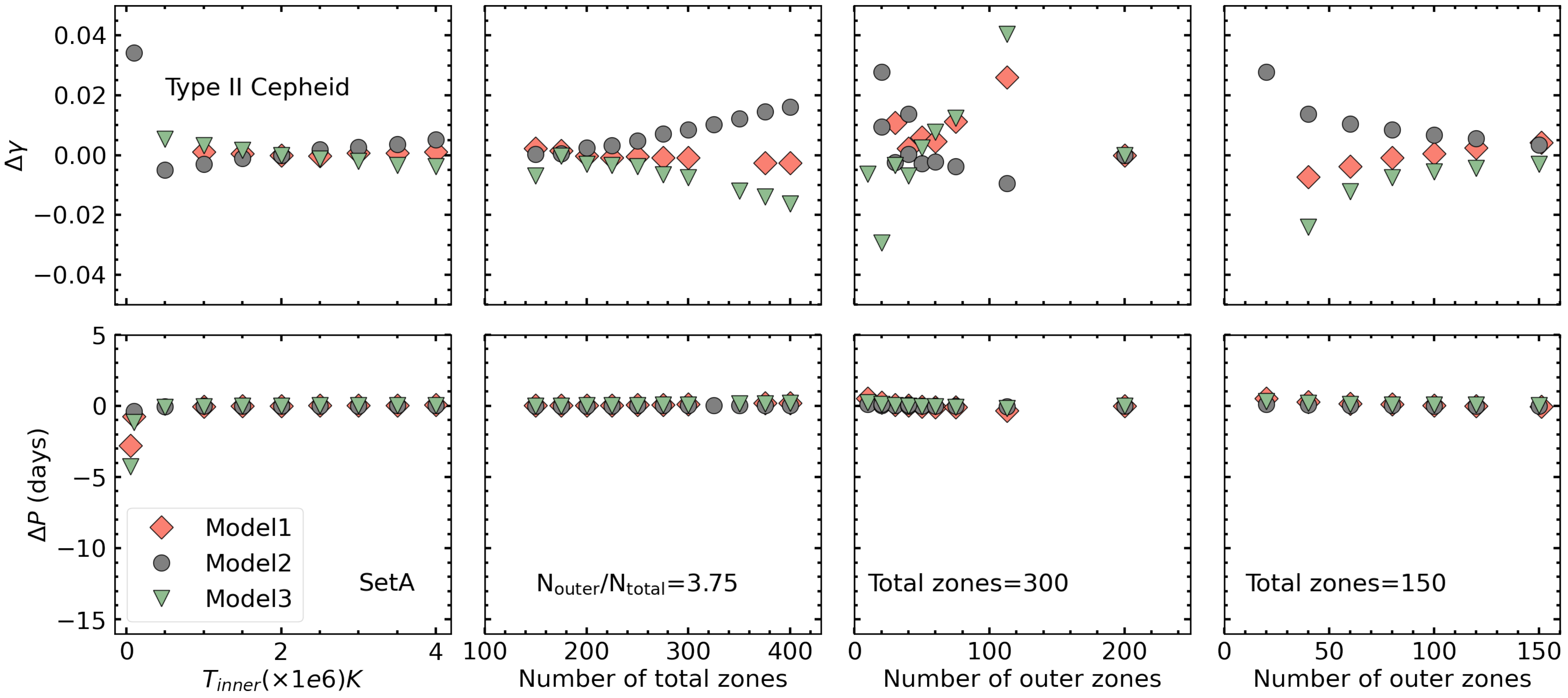}}\\
\vspace{-0.02\linewidth}
\end{tabular}
\caption{The difference in growth rates and periods is plotted against the different input parameters taken for set A. Models ($1-3$) for fundamental
and first-overtone mode Classical Cepheids are chosen randomly from the Classical Cepheids input grid, while models ($1-3$) for Type II Cepheids are selected from Type II Cepheids input grid.}
\label{fig:nmp_setA}
\end{figure*}

%\section{Numerical Parameter Test}
\label{sec:test}
To see the effect of total number of zones ($N_{\rm total}$), outer zones ($N_{\rm outer}$), inner boundary temperature  
$(T_{\rm{inner}})$ and the ratio of outer zones to total zones ($\frac{N_{\rm{outer}}}{N_{\rm total}}$) on growth rates and periods, we randomly choose three models with different chemical composition from each of the CC and T2C input grid.
Then we compute these models using \textsc{MESA-RSP} with different choices of $N_{\rm total}$, $N_{\rm outer}$, 
$T_{\rm{inner}}$ and $\frac{N_{\rm{outer}}}{N_{\rm total}}$. Each parameter is varied at a time while keeping the other
parameters fixed with their default values as given in \citet{paxt19}. The input stellar parameters ($Z, X, M, L, T_{\rm eff}$) are listed in Table~\ref{table:nmp_param}. We have also provided the complete input parameter grids constructed for this calculation in \url{https://github.com/mami-deka/LNA_CMD}.
The result obtained for set A is 
displayed in Fig.~\ref{fig:nmp_setA}. However, it can be seen that the choices of different sets of parameters do not affect the growth rates and the periods much, except for cases where a $T_{\rm{inner}}$ value falls below $0.5\times10^{6}$~[K]. This is true for all four turbulent convection
parameter sets. Hence, we have kept these parameters the same as given in \citet{paxt19}. 

To see the effect of the total number of Lagrangian mass cells on the instability edges, we have also run the CCs grid with a new combination of Lagrangian mass cells $N_{\rm total}=300$ with $N_{\rm outer}=80$ cells using set A. 
However, this change in the zone numbers does not affect the boundaries of instability strips significantly. 

\begin{table*}
\centering
\caption{The stellar parameters used for investigating the sensitivity of the numerical parameters
($N_{\rm total}, N_{\rm outer}, T_{\rm{inner}}$ and $\frac{N_{\rm{outer}}}{N_{\rm total}}$) on the growth rates and linear periods. These models are selected randomly from the original input grids. }
\label{table:nmp_param}
\begin{tabular}{lccccccr} % four columns, alignment for each
\hline
Class of variables & $Z$ & $X$ & $M/\rm{M_{\sun}}$ & $L/\rm{L_{\sun}}$ & $T_{\rm eff}$~(K) & Model reference in Fig.~\ref{fig:nmp_setA}  \\ \hline
CCs (for both FU and FO mode) & $0.013$ & $0.71847$ & $4.089$ & $873.788$ & $6032.812$ & Model1\\
    & $0.00609$ & $0.74541$ & $7.799$ & $9154.214$ & $4855.664$ & Model2 \\
    & $0.0024$ & $0.74541$ & $10.865$ & $22794.218$ & $5663.159$ & Model3\\
T2Cs & $0.013$ & $0.71847$ & $0.381$ & $289.375$ & $5212.500$ & Model1 \\
    & $0.00834$ & $0.73032$ & $0.526$ & $241.357$ & $6028.711$ & Model2 \\
    & $0.00424$ & $0.74073$ & $0.948$ & $752.454$ & $5248.388$ & Model3\\
\hline
\end{tabular}
\end{table*}

\section{The CMDs with the Observed data}
\begin{figure*}
\includegraphics[width=0.9\textwidth,keepaspectratio]{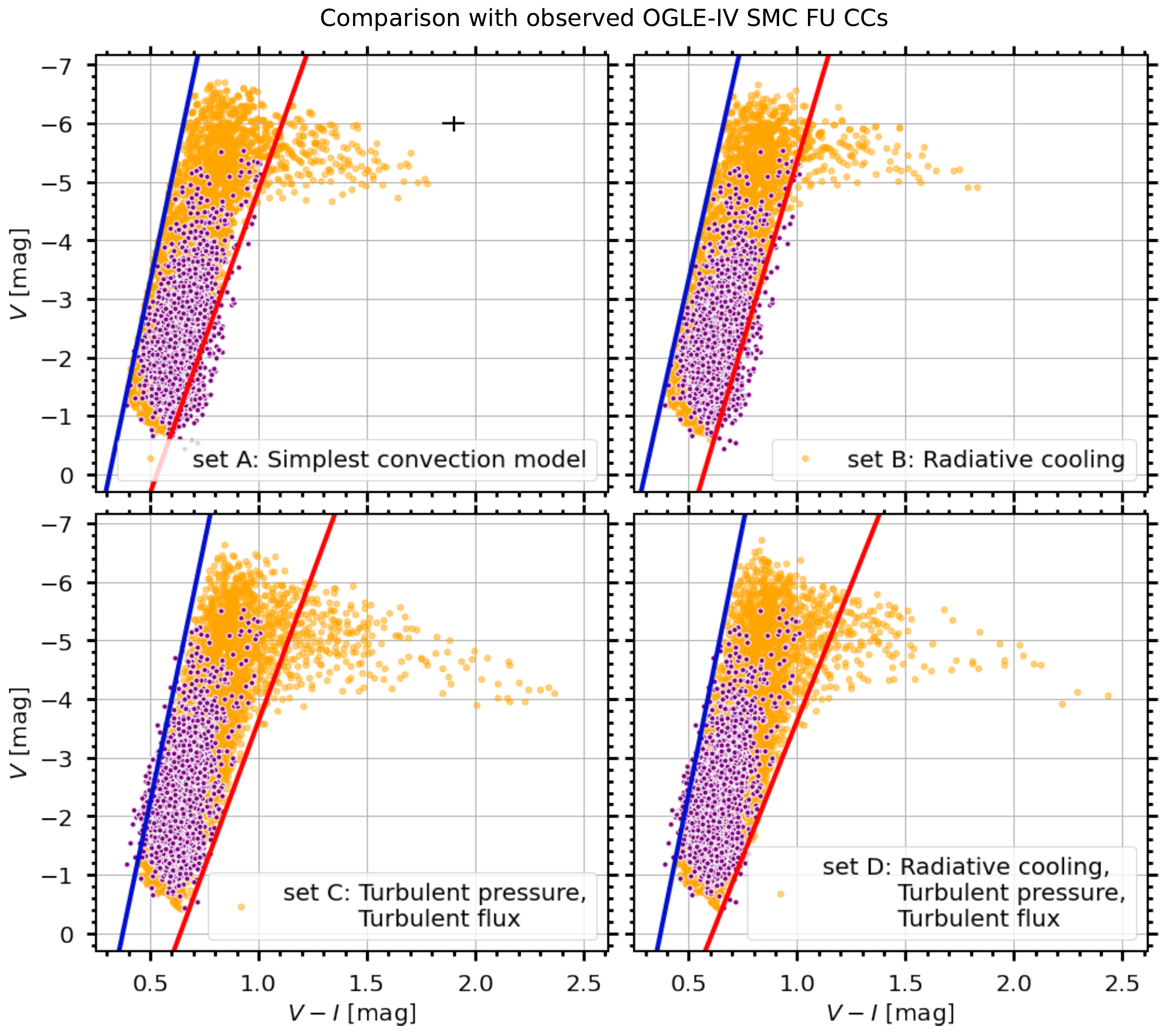}
\caption{Same as Fig.\ref{fig:cmd_cep_FU_lmc_th}, here with observed SMC FU CCs.}
\label{fig:cmd_cep_FU_smc_th}
\end{figure*}

\begin{figure*}
\includegraphics[width=0.9\textwidth,keepaspectratio]{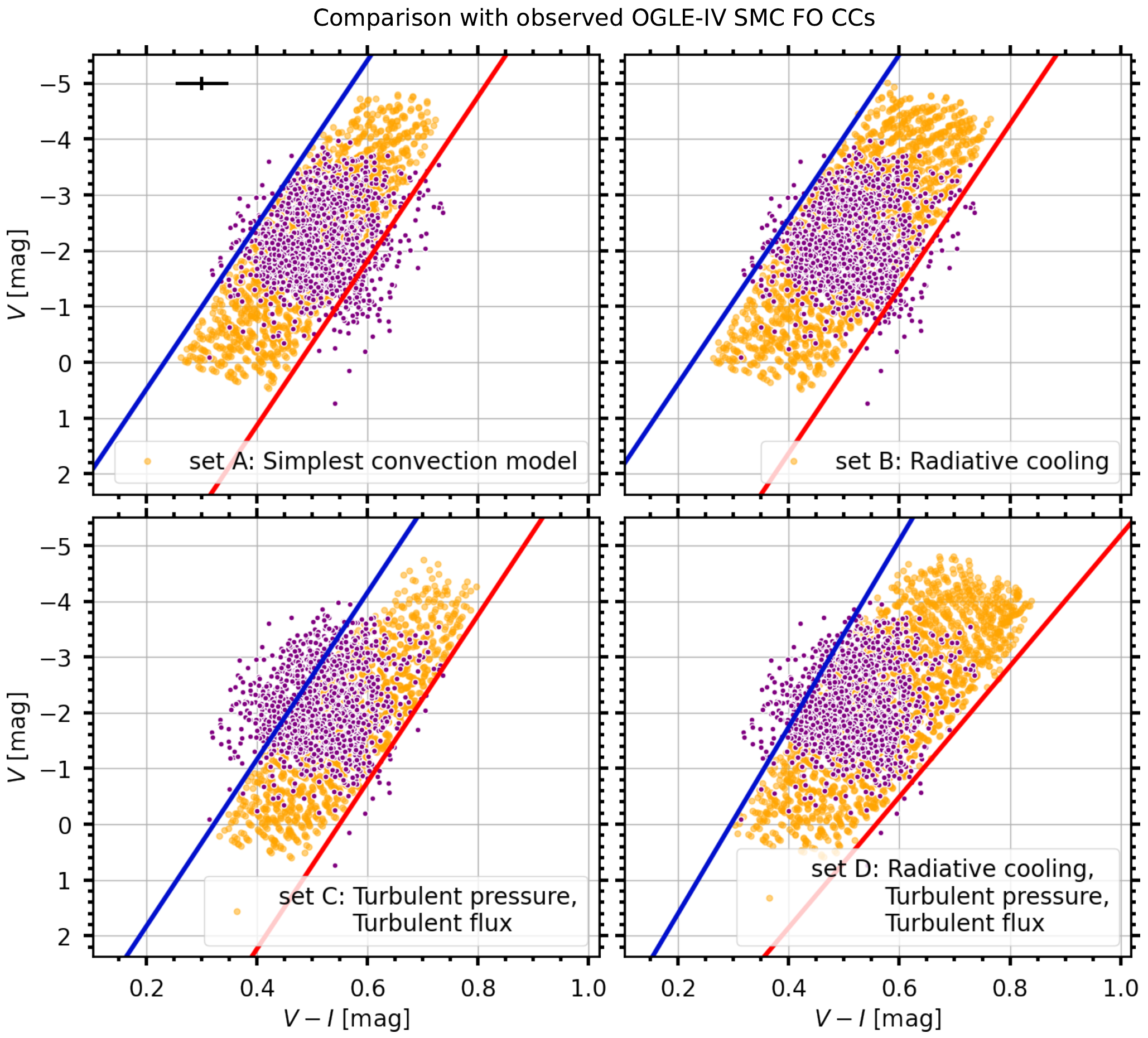}
\caption{Same as Fig.\ref{fig:cmd_cep_FO_lmc_th}, here with observed SMC FO CCs.}
\label{fig:cmd_cep_FO_smc_th}
\end{figure*}

\begin{figure*}
\centering
\includegraphics[width=0.9\textwidth,keepaspectratio]{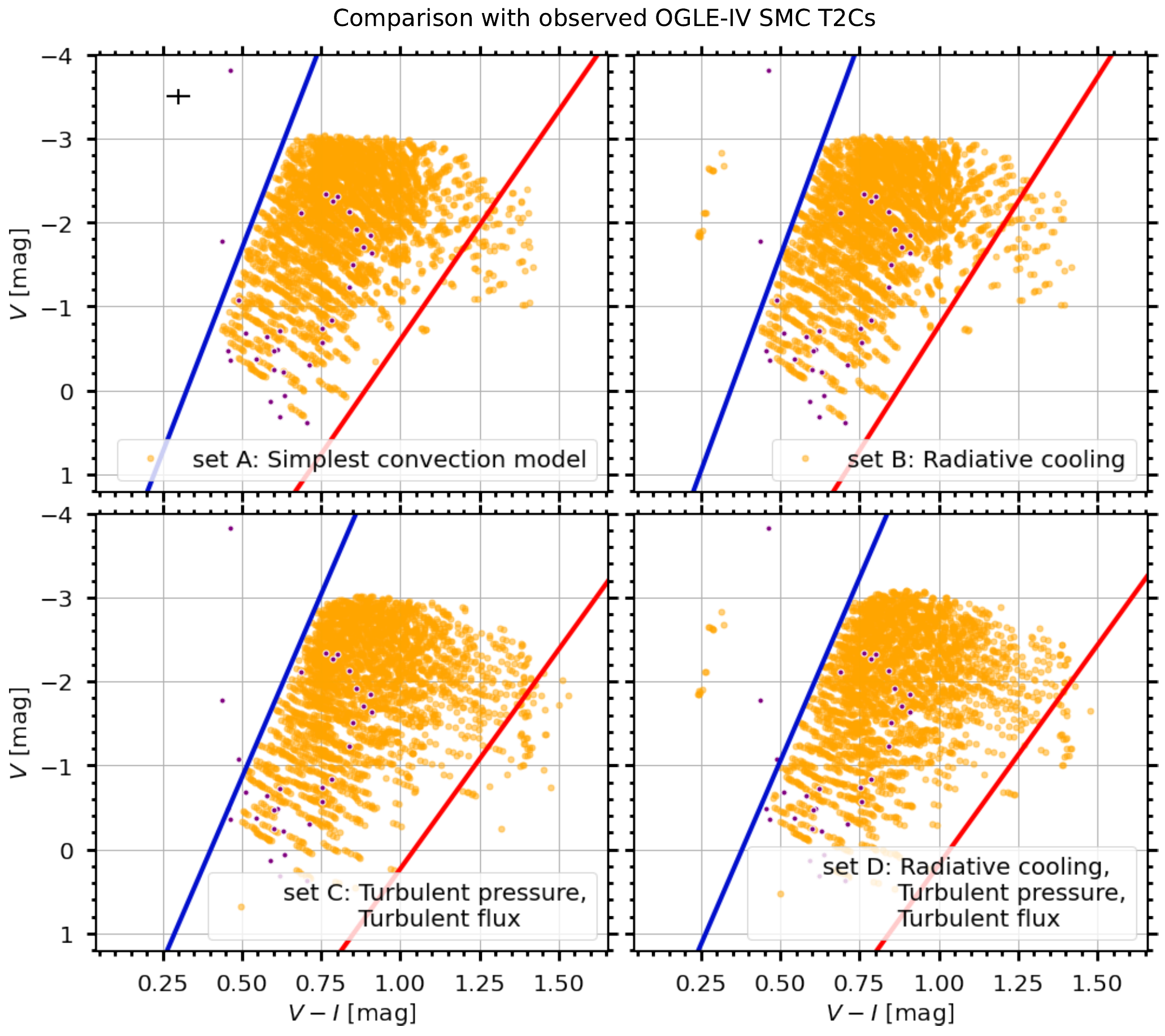}
\caption{Same as Fig.\ref{fig:cmd_cep_FU_lmc_th}, here with observed SMC T2Cs.}
\label{fig:cmd_t2cep_smc_th}
\end{figure*}

Figs.~\ref{fig:cmd_cep_FU_smc_th},~\ref{fig:cmd_cep_FO_smc_th} and ~\ref{fig:cmd_t2cep_smc_th} show the theoretical IS edges with the observed SMC
data. 
% Fig.~\ref{fig:cep_IS} shows the observed data both from LMC and SMC with the theoretical IS edges obtained using different sets of turbulent 
% convection parameters in a single plot. Each panel in these figures corresponds to FU CCs, FO CCs, and T2CS, respectively.

% \begin{figure*}
%   \begin{minipage}{0.5\textwidth}
%     \centering
%     \includegraphics[width=0.9\linewidth]{Figures/cep_instability_edge_cmd_FU.pdf}
%     %\caption{Caption for Figure 1}
%     \label{fig:figure1}
%   \end{minipage}%
%   \begin{minipage}{0.5\textwidth}
%     \centering
%     \includegraphics[width=0.9\linewidth]{Figures/cep_instability_edge_cmd_FO.pdf}
%     %\caption{Caption for Figure 2}
%     \label{fig:figure2}
%   \end{minipage}

%   \vspace{0.5cm}

%   \begin{minipage}{\textwidth}
%     \centering
%     \includegraphics[width=0.45\linewidth]{Figures/t2cep_instability_edge_cmd_FU.pdf}
%     %\caption{Caption for Figure 3}
%     \label{fig:figure3}
%   \end{minipage}
  
%   \caption{Theoretical IS edges in terms of CMD with the four convection sets A, B, C, and D. Red and blue colors represent the red and blue edges, respectively. Left panel: FU mode CCs, middle panel: FO mode CCs, and right panel: T2Cs.}
%   \label{fig:cep_IS}
% \end{figure*}

\section{Additional tables and figures showing the properties of  the pulsation models}
We have listed the slopes and intercepts of the predicted IS edges in HRD as a function of different convection sets in Table~\ref{table:hrd_IS}.
Table~\ref{table:metal_IS} shows the same but as a function of metallicity across different convection sets. 

\begin{center}
% \documentclass{article}
% \usepackage{multirow}
% \usepackage{threeparttable}

% \begin{document}

\begin{table*}
    \begin{threeparttable}
        \centering
        \caption{Slopes and intercepts of IS edges for FU CCs, FO CCs, and T2Cs in the HRD.}
        \label{table:hrd_IS}
        \begin{tabular}{lccccccccr} % seven columns, alignment for each
        \hline
        \multirow{4}{*}{Class of Variables} & Convection Sets & \multicolumn{4}{c|}{IS edges ($\log(L/\rm L_{\odot})=c\log{T_{\rm eff}+d}$)} & \multicolumn{2}{c|}{Width of the IS in $T_{\rm eff}$ (in K)\tnote{*}} \\ \cline{3-8}
                           &                 &\multicolumn{2}{c|}{Blue edge} & \multicolumn{2}{c|}{Red edge} & At minimum  & At maximum \\  \cline{3-6} 
                           &                 & Slope & Intercept & Slope & Intercept & $L/\rm L_{\odot}$ & $L/\rm L_{\odot}$   \\ \hline
          FU CCs       &A              & $-25.239$&$ 99.299$&$-17.039$&$ 66.748$ & 892 & 1210 \\
                           &B              & $-27.519$&$107.826$&$-17.823$&$ 69.518$ & 906 & 1248 \\
                           &C              & $-26.184$&$102.449$&$-15.111$&$ 58.943$ & 1050 & 1510 \\
                           &D              & $-28.184$&$110.099$&$-15.111$&$ 58.823$ & 1149 & 1660\\ \hline
           
          FO CCs       &A              &$-23.506$&$ 92.850$&$-23.506$&$ 91.600$ & 878 & 707 \\ 
                   &B         &$-23.506$&$ 92.800$&$-23.506$&$ 91.400$ & 971 & 783 \\
                   &C              &$-23.506$&$ 92.400$&$-23.506$&$ 91.150$ & 840 & 677 \\
                   &D              &$-23.506$&$ 92.550$&$-23.506$&$ 90.900$ & 1104 & 890 \\ \hline
              T2Cs         &A              &$-16.522$&$65.634$&$ -6.463$&$26.145$ & 1371 & 2362\\
                           &B              &$-16.922$&$67.184 $&$ -6.463$&$26.145$ & 1374 & 2390 \\
                       &C              &$-16.922$&$ 66.934$&$ -5.663$&$22.895$ & 1656 & 2735 \\
                   &D         &$-16.922$&$66.964$&$ -5.663$&$22.975 $ & 1504 & 2651 \\ \hline
        \end{tabular}
        \begin{tablenotes}
            \item[*] Note: The strip width is calculated at minimum and maximum luminosity of the pulsating models available in each grid.
        \end{tablenotes}
    \end{threeparttable}
\end{table*}

% \end{document}

\end{center}

\begin{center}
\begin{table*}
        \centering
        \caption{Slopes and intercepts of IS edges for FU CCs, FO CCs and T2Cs at different metalicities in the CMD.}
        \label{table:metal_IS}
        \begin{tabular}{|c|c|c|c|c|c|c|} % four columns, alignment for each
        \hline
		Class of Variable & Metal Fraction ($Z$) & Convection Set & \multicolumn{4}{c|}{IS edges  $(V=a(V-I)+b)$}  \\ \cline{4-7}
			   & &                 &\multicolumn{2}{c|}{Blue edge} & \multicolumn{2}{c|}{Red edge} \\  \cline{4-7} 
			   & &                 & Slope & Intercept & Slope & Intercept \\ \hline
		FU CCs     &$0.013$    &A              &$-12.880$&$3.594$&$-10.368$&$5.506$\\
			   &           &B		 &$-16.880$&$5.094$&$-10.388$&$5.553$ \\
			   &           &C              &$-11.880$&$4.194$&$-8.588$&$5.253$ \\
			   &           &D              &$-13.280$&$4.994$&$-8.588$&$5.153$ \\ \cline{2-7}
	   	           &$0.00609$  &A              &$-15.880$&$4.594$&$-10.288$&$5.253$\\
			   &           &B		 &$-16.380$&$5.094$&$-10.288$&$5.453$ \\
			   &           &C              &$-11.880$&$3.594$&$-8.588$&$5.153$ \\
			   &           &D              &$-15.980$&$5.794$&$-10.288$&$6.153$ \\ \cline{2-7}
			   &$0.0024$   &A              &$-17.568$&$5.493$&$-10.588$&$5.453$ \\
			   &   	       &B		 &$-16.522$&$4.941$&$-10.588$&$5.453$ \\
			   &   	       &C              &$-13.880$&$4.394$&$-8.588$&$4.953$ \\
			   &           &D              &$-16.380$&$5.794$&$-10.288$&$5.953$ \\ \hline

		FO CCs     &$0.013$    &A         &$-10.452$&$3.832$&$-14.752$&$6.832$ \\
			   &           &B              &$-10.452$&$3.232$&$-14.752$&$7.332$ \\
			   &           &C              &$-11.370$&$5.332$&$-14.752$&$7.932$ \\
			   &           &D		         &$-11.752$&$3.232$&$-12.752$&$7.232$ \\ \cline{2-7}
			   &$0.00609$  &A              &$-10.452$&$3.632$&$-14.752$&$6.832$ \\
			   &           &B              &$-13.792$&$3.532$&$-14.752$&$7.332$ \\
			   &           &C              &$-14.752$&$4.932$&$-14.752$&$7.932$ \\
			   &           &D		 &$-14.752$&$4.232$&$-12.752$&$7.032$ \\ \cline{2-7}
			   &$0.0024$   &A              &$-11.370$&$3.432$&$-14.752$&$7.032$ \\
			   &           &B              &$-14.752$&$3.432$&$-14.752$&$7.132$ \\
			   &           &C              &$-14.752$&$4.532$&$-14.752$&$8.032$ \\
			   &           &D		 &$-14.752$&$4.132$&$-12.752$&$6.832$ \\ \hline

		T2Cs       &$0.013$   &A           &$-10.452$&$3.796$&$-4.982$&$4.093$ \\
			   &            &B              &$-10.452$&$3.596$&$-4.482$&$3.793$ \\
			   &            &C              &$-10.452$&$4.596$&$-3.582$&$3.993$ \\
			   &            &D		  &$-8.952$&$3.496$&$-3.982$&$4.293$ \\ \cline{2-7}
                           &$0.00834$   &A &$-10.452$&$3.596$&$-4.982$&$4.093$ \\
			   &            &B              &$-11.852$&$4.396$&$-4.982$&$4.193$ \\
			   &            &C              &$-9.152$&$3.796$&$-4.982$&$5.093$ \\
			   &            &D		  &$-8.852$&$3.495$&$-5.482$&$5.193$ \\ \cline{2-7}
	                   &$0.00424$   &A   &$-11.370$&$4.497$&$-4.519$&$3.741$ \\
			   &            &B              &$-11.370$&$4.297$&$-4.919$&$4.141$ \\
			   &            &C              &$-11.370$&$4.297$&$-5.119$&$5.041$ \\
			   &            &D		  &$-9.970$&$4.197$&$-7.119$&$6.341$ \\ \cline{2-7}
	                   &$0.00135$   &A    &$-10.792$&$3.988$&$-5.412$&$4.199$ \\
			   &            &B              &$-13.292$&$4.988$&$-5.612$&$4.599$ \\
			   &            &C              &$-7.792$&$2.988$&$-5.412$&$5.199$ \\
			   &            &D		  &$-9.792$&$3.988$&$-5.412$&$4.799$ \\ \cline{2-7}
	                   &$0.00061$   &A     &$-13.792$&$5.488$&$-5.812$&$4.599$ \\
			   &            &B              &$-13.792$&$5.488$&$-5.812$&$4.599$ \\
			   &            &C              &$-7.792$&$2.988$&$-5.812$&$5.399$ \\
			   &            &D		  &$-9.792$&$3.988$&$-5.812$&$5.099$ \\ \cline{2-7}
			   &$0.00043$   &A              &$-13.792$&$5.488$&$-5.812$&$4.599$ \\
			   &            &B              &$-13.792$&$5.488$&$-5.812$&$4.599$ \\
			   &            &C              &$-7.792$&$2.988$&$-5.812$&$4.999$ \\
			   &            &D		  &$-9.792$&$3.988$&$-5.812$&$4.999$ \\ \cline{2-7}
	                   &$0.00014$   &A      &$-15.292$&$6.288$&$-5.812$&$4.799$ \\
			   &            &B              &$-15.292$&$6.288$&$-6.212$&$4.999$ \\
			   &            &C              &$-8.792$&$3.688$&$-5.812$&$5.099$ \\
			   &            &D		  &$-9.792$&$3.988$&$-5.812$&$5.099$ \\ \hline

        \end{tabular}
\end{table*}

\end{center}

% \begin{figure*}
% \centering
% \includegraphics[width=1.0\textwidth,keepaspectratio]{Figures/tc_z_cbar.pdf}
% \caption{Plot of mean $T_{\rm eff}$ for FU CCs, FO CCs and T2Cs obtained from Gaussian fitting as a function of
% different metallicities and different sets of turbulent convection parameters.}
% \label{fig:quant_temp1}
% \end{figure*}

% \begin{center}
% \input{quant_temp.tex}
% \end{center}
\begin{figure*}
\centering
\includegraphics[width=1.0\textwidth,keepaspectratio]{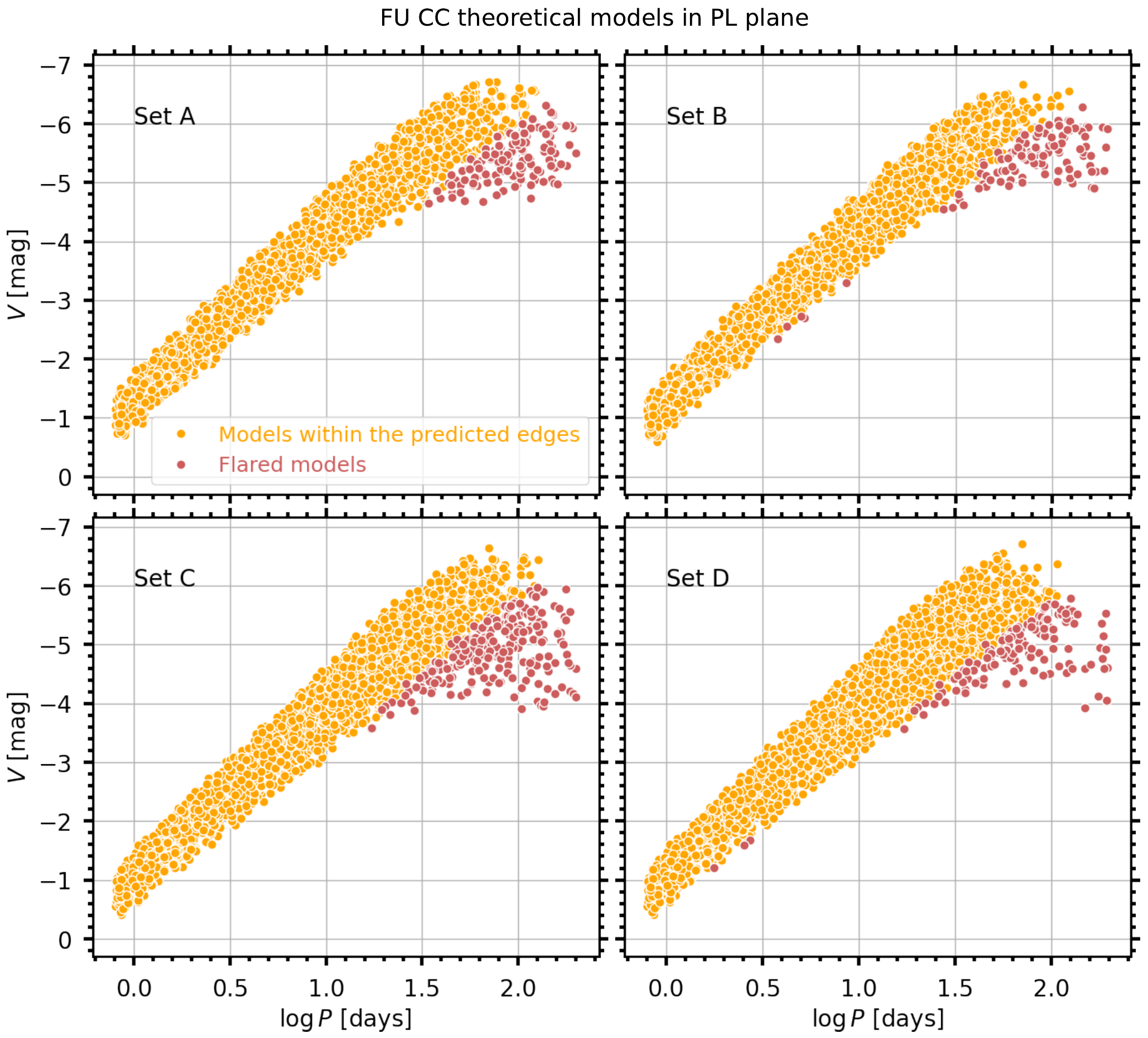}
\caption{The plot shows the theoretical pulsating models of FU classical Cepheids in period-luminosity plane. The orange dots are the models within our predicted IS edges while the red dots represent the flared models.}
\label{fig:PL_cep}
\end{figure*}

% Don't change these lines
\bsp	% typesetting comment
\label{lastpage}
\end{document}